\documentclass[12pt,preprint]{aastex}

\newcommand{\ppm}{{\rm \; ppm}}

\newcommand{\dex}{{\rm \; dex}}

\newcommand{\logMgH}{{\log{{\rm(Mg/H)}}}}
\newcommand{\logMgIIH}{{\log{{\rm(Mg \; II/H)}}}}

\newcommand{\Htot}{{\rm H_{tot}}}

\newcommand{\NHI}{{N{\rm(H \; I)}}}

\newcommand{\NHmol}{{N\rm(H_{2})}}

\newcommand{\logMgII}{{\log{N{\rm(Mg \; II)}}}}

\newcommand{\logHI}{{\log{\NHI}}}
\newcommand{\logHmol}{{\log{\NHmol}}}
\newcommand{\logHtot}{{\log{\NHtot}}}

\newcommand{\NHtot}{{N{\rm(H_{tot})}}}
\newcommand{\Hmol}{{\rm H_{2}}}

\newcommand{\nHavg}{{< \! \! \rm{n_{H}} \! \! >}}

\newcommand{\lognHavg}{{\log{\nHavg}}}

\newcommand{\fHmol}{{f\rm(H_{2})}}
\newcommand{\eqw}{{W_{\lambda}}}
\newcommand{\av}{{A_V}}

\newcommand{\ebv}{{E_{B-V}}}

\newcommand{\avdist}{{\av / r}}
\newcommand{\ebvdist}{{\ebv / r}}
\newcommand{\rv}{{R_V}}

\shorttitle{}

\shortauthors{Jensen and Snow}

\begin{document}

\title{THE VARIATION OF MAGNESIUM DEPLETION WITH LINE OF SIGHT CONDITIONS}

\author{Adam G. Jensen and Theodore P. Snow}

\affil{Center for Astrophysics and Space Astronomy}
\affil{University of Colorado at Boulder, Campus Box 389}
\affil{Boulder, CO 80309-0389}

\email{Adam.Jensen@colorado.edu, tsnow@casa.colorado.edu}

\begin{abstract}
In this paper we report on the gas-phase abundance of singly-ionized magnesium (Mg II) in 44 lines of sight, using data from the {\it Hubble Space Telescope} ({\it HST}).  We measure Mg II column densities by analyzing medium- and high-resolution archival STIS spectra of the 1240 \AA{} doublet of Mg II.  We find that Mg II depletion is correlated with many line of sight parameters (e.g.~$\fHmol$, $E_{B-V}$, $\ebvdist$, $A_V$, and $\avdist$) in addition to the well-known correlation with $\nHavg$.  These parameters should be more directly related to dust content and thus have more physical significance with regard to the depletion of elements such as magnesium.  We examine the significance of these additional correlations as compared to the known correlation between Mg II depletion and $\nHavg$.  While none of the correlations are better predictors of Mg II depletion than $\nHavg$, some are statistically significant even assuming fixed $\nHavg$.  We discuss the ranges over which these correlations are valid, their strength at fixed $\nHavg$, and physical interpretations.
\end{abstract}


\section{INTRODUCTION AND BACKGROUND}
\label{s:MgII_intro}
Magnesium is both a relatively abundant element in the Galaxy and an important component in most interstellar dust models.  Mg I has an ionization potential of only 7.65 eV, while Mg II has an ionization potential of 15.04 eV.  In H I regions, the dominant form of gas-phase interstellar magnesium should be Mg II.  In H II regions, magnesium should be found primarily in the form of Mg III.  Gas-phase Mg I is rare, even in $\Hmol$ regions.  As is the case for other elements such as silicon and iron, the average gas-phase abundance of magnesium is much smaller than the assumed overall cosmic abundance of magnesium, implying that the majority of interstellar magnesium is tied up in dust.  Therefore, variations in the gas-phase magnesium abundance are only capable of having minor effects on grain composition.  Nevertheless, observed variations in the gas-phase magnesium abundance may still shed light on the physical conditions of interstellar clouds.



Two major {\it Copernicus} surveys that included measurements of Mg II abundances and depletions were \citet{Murray} and \citet{Jenkins1986}.  Both of these studies confirmed that magnesium depletion increases with increased average hydrogen volume density, $\nHavg=\NHtot/r$ (where $r$ is the line-of-sight pathlength), in the line of sight.  \citet{Jenkins1986} also found several correlations between the depletions of magnesium and other elements, which they cited as secondary to the correlation between those depletions and $\nHavg$.  In both studies, the depletion of magnesium was not strongly correlated to other line of sight parameters, such as $\ebv$ and the magnitude of the 2175 \AA{} extinction ``bump''.  We also note that there is a systematic difference in the absolute values of the abundances and depletions between these studies and more recent studies that is fairly substantial.  This is due to a difference in the assumed $f$-values of the 1240 \AA{} doublet of magnesium.  We comment on our chosen $f$-values in \S \ref{ss:MgII_lines}.

More recently, \citet{Cartledge2006} also examined the abundances and depletions of Mg II and several other elements in the interstellar medium using STIS data.  (Hereafter, this paper will be referred to as CLMS, for the initials of the authors.)  Similar to the {\it Copernicus} studies, CLMS concluded that the line of sight parameter with the clearest connection to elemental depletions is the average hydrogen volume density, $\nHavg$.  CLMS also explored potential correlations between depletions and other line of sight parameters such as the molecular fraction of hydrogen, $\fHmol$; selective extinction, $\ebv$; and selective extinction divided by line-of-sight pathlength, $\ebvdist$.  CLMS concluded that $\nHavg$ was the parameter that best identified warm vs.~cold clouds, and that no other parameters produced correlations with magnesium depletion that were both as strong and with as little scatter.

We began this study before CLMS was published.  In spite of the similarities between that study and this one, we have proceeded with this study to provide an independent analysis of Mg II, and also because we have analyzed potential trends with respect to parameters not analyzed in CLMS, e.g.~$\av$ and $\rv$.  The Mg II column densities of 11 out of the 44 lines of sight in our sample have also been analyzed previously by CLMS.  We still report our results in this paper for two main reasons:  (1) these lines of sight provide a basis of comparison for the methods of this paper and those of CLMS and (2) these lines of sight can be analyzed with respect to the aforementioned line of sight parameters not analyzed in CLMS.  A comparison of the column density measurements for these common lines of sight is found in \S \ref{ss:MgII_coldensities}.

In \S \ref{s:MgII_obsdata} we discuss our observations and data reduction, including comments on the 1240 \AA{} doublet and our derivation of column densities and abundances.  In \S \ref{s:results} we discuss our results, including observed correlations and a review of the Galactic abundance of magnesium.  In \S \ref{s:MgII_summary} we summarize our findings.

\section{OBSERVATIONS AND DATA REDUCTION}
\label{s:MgII_obsdata}

\subsection{The Mg II Doublet at $\lambda\lambda$1239,1240}
\label{ss:MgII_lines}
\citet{Morton2003}, summarizing several sources, cites only four major, ground-state doublets of Mg II---a strong doublet with transitions at 2795.5 \AA{} and 2802.7 \AA{}, a much weaker doublet at 1239.9 \AA{} and 1240.4 \AA{}, a doublet at 1026.0 \AA{} and 1026.1 \AA{} similar in strength to the 1240 \AA{} doublet, and an even weaker doublet at 946.7 \AA{} and 946.8 \AA{}.  STIS data sets that cover the wavelength of strong doublet near 2800 \AA{} exist for only one of our lines of sight (HD 93205); if more data did exist, the lines would be saturated in most cases, though each line might show damping wings for a more certain derivation of column density in a few cases.  The doublets at 946 \AA{} and 1026 \AA{}, in addition to being self-blended in {\it FUSE} data, are both wiped out by strong hydrogen absorption in the high-column density lines of sight in this sample ($\Hmol$ wipes out the 946 \AA{} doublet and H I Lyman-$\beta$ absorption wipes out the 1026 \AA{} doublet).

Thus, the two lines of the 1240 \AA{} doublet are the only absorption lines of use.  \citet{Morton2003} notes that there are converging theoretical values for the combined magnitude of the $f$-values of this doublet, though there is some variation between different authors in the calculated ratio of the $f$-values.  The studies under discussion in \citet{Morton2003} include \citet{Fleming1998}, \citet{GFF1999}, \citet{TF1999}, and \citet{Majumder2002}.  \citet{Morton2003} ultimately opts for the values by \citet{TF1999}.  These sources calculate a combined $f$-value of the doublet that is consistent to within 0.13$\dex$.  In keeping with \citet{Morton2003}, we elect to use the \citet{TF1999} $f$-values of $f_{1239.9}=6.32\times10^{-4}$ and $f_{1240.4}=3.56\times10^{-4}$.  CLMS also opted to use these $f$-values; thus, $f$-values are not a source of systematic error between this study and CLMS.  \citet{Morton2003} notes that despite consistency in the combined $f$-value of the doublet, there is some discrepancy in the $f$-value ratio between the two lines.  However, the \citet{TF1999} $f$-values represent a ratio (1.78) that is consistent with empirical values by \citet{Fitzpatrick1997} and \citet{SFH2000} and the theoretical calculations of \citet{Majumder2002}; these sources quote ratios between 1.74 and 1.82.  Note, however, that \citet{GFF1999} found a ratio of 2.54, while \citet{Fleming1998} only calculated the combined $f$-value of the doublet and not of the individual lines.

The damping constants of these lines are not reported in \citet{Morton2003}, but in all cases the lines are much too weak for damping constants to have a significant effect.  We use this doublet exclusively for the determination of column densities.  We describe our fitting methods below (\S \ref{ss:MgII_coldensities}), but first describe the STIS data and relevant procedures used to observe the lines.

\subsection{{\it HST} Data}
\label{ss:MgII_HSTdata}
Archival STIS data are available for all 44 of the lines of sight in this study.  Lines of sight were initially chosen from several literature sources in attempts to examine abundances of silicon and iron.  Basic line of sight parameters (e.g.~spectral types and Galactic coordinates of the background stars) are given in Table \ref{MgII_stellardata}.  A summary of the STIS data sets used is given in Table \ref{MgII_obstable}.  Hydrogen column densities and related parameters (e.g.~$\fHmol$) are given in Table \ref{MgII_hydrogentable}, while extinction and reddening parameters $\ebv$, $\av$, and $\rv$ are given in Table \ref{MgII_reddeningtable}.

We used ``on-the-fly'' calibrated \citep{Micol} data.  For observations with the E140H grating, the 1240 \AA{} line is found in two echelle orders.  We coadded these echelle orders and multiple observations where available.  We performed the same fitting routines described below (\S \ref{ss:MgII_coldensities}) in test cases on the individual echelle orders and observations and found consistency between those results and our results for the coadded spectra.

We assumed empirically measured point-spread functions (S. V. Penton, private communication) in analyzing data that utilize the E140H and E140M gratings and various apertures.  We have noted the use of this PSF in our previous work \citep{JensenNI, JensenFeII}.  This empirical PSF is not Gaussian.

\subsection{Column Density Measurements and Errors}
\label{ss:MgII_coldensities}
Measurements of Mg II column densities are derived solely through measurements of the 1240 \AA{} doublet.  Assuming the $f$-values given in \S \ref{ss:MgII_lines} and the PSF described in \S \ref{ss:MgII_HSTdata}, we fit Voigt profiles simultaneously to the two lines of the doublet.  The Voigt profile is the most general absorption profile; it is inherently Gaussian for unsaturated weak lines, and inherently damped for very strong lines.  The observed profile is the convolution of the inherent profile and the PSF.  When the lines of the doublet are asymmetric or multiple velocity components are clearly resolved, we fit each individual component with a Voigt profile.

We fit these multiple components without any {\it a priori} assumptions on the overall velocity structure of the line of sight.  This involves ``$\chi$-by-eye'' to first order, then comparing the reduced $\chi^2$ of fits with a different number of components.  Though high-resolution optical data on the velocity structure exist for some of these lines of sight \citep[e.g.][]{Pan,Welty2001,Welty2003}, the ionization potential of the species studied (K I and Ca I) is smaller than the ionization potential of Mg I (7.6 eV) and much smaller than the ionization potential of Mg II (15.1 eV), indicating that those elements and Mg II are unlikely to have high spatial coincidence.  Therefore, fitting observed resolved components and asymmetries is no less---and mostly likely better---justified than assuming the velocity structure of one of these other elements.

\begin{figure}[t!]
\begin{center}
\epsscale{1.00}
\plotone{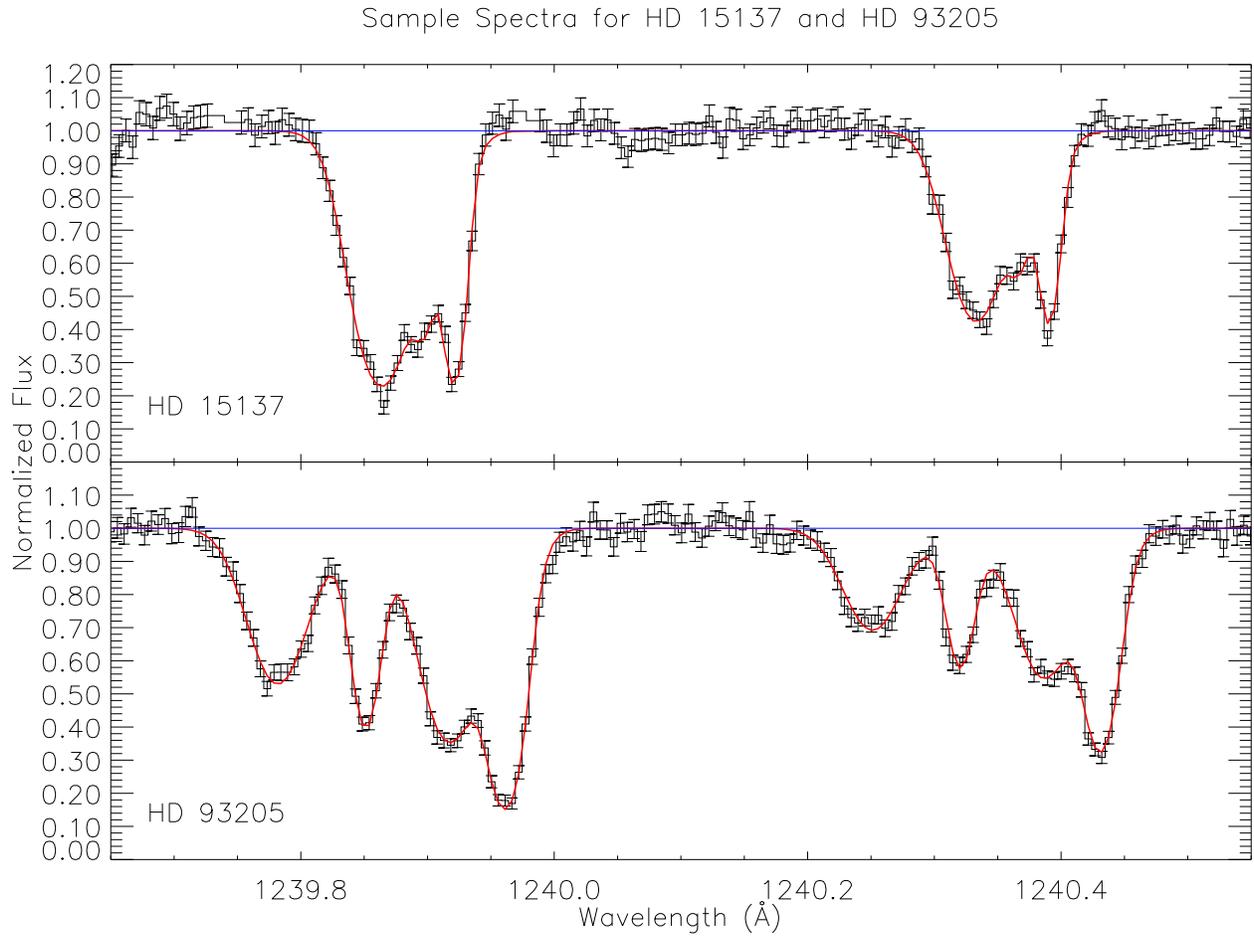}
\end{center}
\caption[Sample of Fits to Mg II 1240 \AA{} Doublet]{Two samples of the fits made by our techniques.  Spectra are shown normalized but without velocity correction.  Error bars for the STIS data are shown, with profile fits plotted in red and the continuum level plotted in blue.  {\bf Top}:---HD 15137.  {\bf Bottom}:---HD 93205.}
\label{fig:specsample}
\end{figure}

Our profile-fitting\footnote{Note that our use of the phrase ``profile fitting'' is different than that of CLMS, in that, as described above, we do not place any {\it a prior} constraints on the components that we identify.}code simultaneously fits the profiles of both doublet lines with a single total line-of-sight column density, and three parameters for each observed velocity component:  $b$-value, velocity offset, and fraction of total column density.  The code outputs errors for each of these parameters, though the errors on total column density are often somewhat small.  To properly account for continuum placement, which should be the dominant source of error, we calculate the equivalent widths ($\eqw$) and errors in $\eqw$ due to a 1-$\sigma$ shift in the continuum (determined by the S/N).  We then assume that the fractional error in column density is the same as the average fractional error in the $\eqw$ of the two lines of the doublet.  This is a slightly more conservative estimate of the column density error than our code otherwise produces.  Two samples of the fit profiles are shown in Figure \ref{fig:specsample}.  Our column density results, along with derived abundances (see \S \ref{ss:hydrogen}) and equivalent widths, are shown in Table \ref{MgII_coldensities}.

Potential systematic errors in our methods could come from the aforementioned uncertainties in the $f$-values, incorrect normalization of the continuum, and concerns of unresolved or misidentified saturation.  Our choice of $f$-values was discussed above (\S \ref{ss:MgII_lines}).  As noted there, while there is some range in the theoretical values summarized by \citet{Morton2003}, two recent empirical studies determine ratios consistent with our chosen ratio of 1.78, and only one of the three theoretical values that \citeauthor{Morton2003} cites is inconsistent with this value.  Because we cannot guarantee that both components are unsaturated in most cases, we cannot independently analyze the ratio of $f$-values.  However, we feel confident that the aforementioned data are converging toward an $f$-value ratio of $\approx1.78$.

Normalizing the continuum is fairly trivial in this spectral region, as this doublet is far from any other major spectral features in most lines of sight.  In a few lines of sight, broad stellar features do appear, but fitting the background spectra is still easily accomplished with the use of multiple order polynomials.

Saturation can pose a problem in two different ways.  First, if there are unresolved saturated components, our fits might determine a column density that is too small.  Second, if we observe a resolved saturated component, and the inherent width of the line is not much broader than the width of the PSF, the fit will be very uncertain, which could result in significant deviations from the true column density.  Both of these concerns should be somewhat alleviated by the fact that our method of simultaneously fitting the profiles of both lines of the doublet constrains saturation as much as is possible given the data.  This is especially true of the latter concern of misidentifying the $b$-value of a resolved, saturated component.

We also note that in the 11 lines of sight where we have overlapping data with CLMS, our methods produce column densities that are generally consistent with their profile fitting and apparent optical depth methods (hereafter PF and AOD, respectively).  A comparison of the common lines of sight is shown in Table \ref{MgII_comparison}.  Of the 22 comparison pairs (our results versus each method for all 11 lines of sight), 19 possess 1-$\sigma$ agreement.  Another two pairs (our measurements versus the AOD measurements for HD 37021 and HD37903) miss 1-$\sigma$ agreement but are consistent within 2-$\sigma$ and agree within 1-$\sigma$ on the PF measurements.  Conversely, our measurement of HD 147888 agrees with the AOD measurement of CLMS, but not with the PF measurement.  CLMS note this line of sight as by far the most discrepant between the two methods in their study ($0.19\dex$).  However, CLMS also discuss strategies for correction to the AOD method by \citet{SavageSembach1991} and \citet{Jenkins1996}, the former of which brings the two methods to within $0.08\dex$.  CLMS elect to adopt their PF results in all cases.

These comparisons allow us to conclude that our methods are consistent with the (uncorrected) AOD method, which itself is typically consistent with the PF method of CLMS.  The discrepancy for HD 147888 is well outside the norm in this regard.  We also note that our equivalent widths (reported in Table \ref{MgII_coldensities}) are consistent with equivalent widths of CLMS to within errors (though note that CLMS only report the equivalent width of the 1239.9 \AA{} line).

\subsection{Hydrogen Column Densities}
\label{ss:hydrogen}
In order to analyze abundances and depletions, we need to have hydrogen column densities.  We define $\NHtot=[\NHI+2\NHmol]$, neglecting contributions from H II regions.  Given the ionization potentials of H I and Mg II (13.6 eV and 15.1 eV), it is reasonable to assume that the vast majority of magnesium is found in the form of Mg I and Mg II in H I and $\Hmol$ regions, with only a minimal amount of Mg II from H II regions contributing to the Mg II column density that we see.

The atomic hydrogen column densities are determined through profile fitting of the Lyman-$\alpha$ line of atomic hydrogen.  The fits come from several sources, summarized in Table \ref{MgII_hydrogentable}.  Three stars are cooler than type B2.5:  HD 27778, HD 91597, and HD 147888.  Therefore, the Lyman-$\alpha$ fits of these lines might be subject to stellar contamination.  However, in all cases, estimations of stellar Lyman-$\alpha$ and/or consistency checks with other methods of H I determination have been employed.  For a full discussion of these, see the references in which the measurements originally appear (\citeauthor{Cartledge2004}~\citeyear{Cartledge2004} for HD 27778 and HD 147888; \citeauthor{DS1994} \citeyear{DS1994} for HD 91597) or consult our previous discussions of these lines of sight \citep{JensenNI, JensenFeII}.

Molecular hydrogen column densities are taken from recent {\it FUSE} surveys and other {\it FUSE} data (primarily J. M. Shull et al., in preparation, but see Table \ref{MgII_hydrogentable} for all references).  In these papers $\NHmol$ is determined for each line of sight by fitting several low-$J$ lines.  Reliable FUV data do not exist for HD 37021 and HD 37061; therefore, in these two lines of sight we cannot report $\Hmol$ column densities or $\fHmol$, and our $\NHtot$ and $\nHavg$ are only measures of atomic hydrogen column density and volume density, respectively.  This is unfortunate, because as discussed in \S \ref{s:results}, these two lines of sight have two of the five largest depletions of any of the lines of sight in this sample.  If $\NHmol$ is large in either of these lines of sight, then they likely have the greatest depletions of all the lines of sight in this sample, perhaps by a significant amount.  However, \citet{Cartledge2001} have argued that $\fHmol$ is likely small in both of these lines of sight.  This is based on observations of little or no Cl I in these lines of sight; the Cl I ionization potential of 12.97 eV implies that $\NHmol$ is similarly small, in turn implying a small value of $\fHmol$.

\section{RESULTS AND DISCUSSION}
\label{s:results}

\subsection{Correlations}
\label{ss:correlations}
Our results for the column densities of Mg II are given in Table \ref{MgII_coldensities}.  The major correlation found by both \citet{Jenkins1986} and CLMS is that the depletion of magnesium and many other elements increases with overall line of sight density, $\nHavg$.  Both interpreted their models in light of the model by \citet{Spitzer1985}.  The \citeauthor{Spitzer1985} model contends that the ISM contains two distinct varieties of cloud (warm and cold), each with a distinct depletion level.  In this model, lines of sight with average densities of a few-tenths of a particle per cm$^{-3}$ are largely sampling the warm ISM, while lines of sight with average densities of a few particles per cm$^{-3}$ are largely sampling the cold ISM.  This explains the observed plateaus of depletion at both low and high densities, with a transition near average densities of $\sim1$ cm$^{-3}$, where lines of sight are sampling both types of clouds.  This model, however, does not take into account the transition to a much different regime of cloud chemistry expected for translucent clouds \citep[for a recent review, see][]{SnowMcCall}.

\citet{Jenkins1986} also found evidence of lesser correlations between the depletion of various elements and other parameters of reddening and extinction, but concluded that these correlations were secondary to the correlation with $\nHavg$.  CLMS examined the depletion of Mg II compared to $\fHmol$, $\ebv$, and $\ebvdist$.  CLMS concluded that only $\ebvdist$ provided a correlation with magnesium depletion that was as at least roughly as significant as the correlation with $\nHavg$, but with increased scatter.

However, in our recent work \citep{JensenFeII}, we explored the possibility of correlations between iron depletion and various other line of sight parameters.  We found very similar correlations between iron depletion and measures of dust density ($\ebvdist$ and $\avdist$) and the molecular fraction of hydrogen ($\fHmol$).  We also examined iron depletion as a function of $\rv$, the ratio of total visual extinction to selective extinction, though a conclusive trend did not present itself.  Using many of the same lines of sight in this paper as in \citet{JensenFeII}, we find the same correlations generally hold with respect to magnesium depletion.  In what follows, we discuss the nature of these correlations and possible interpretations.

\subsubsection{Correlations with Hydrogen}
\label{sss:hydrogen_correlations}
\begin{figure}[t!]
\begin{center}
\epsscale{1.00}
\plotone{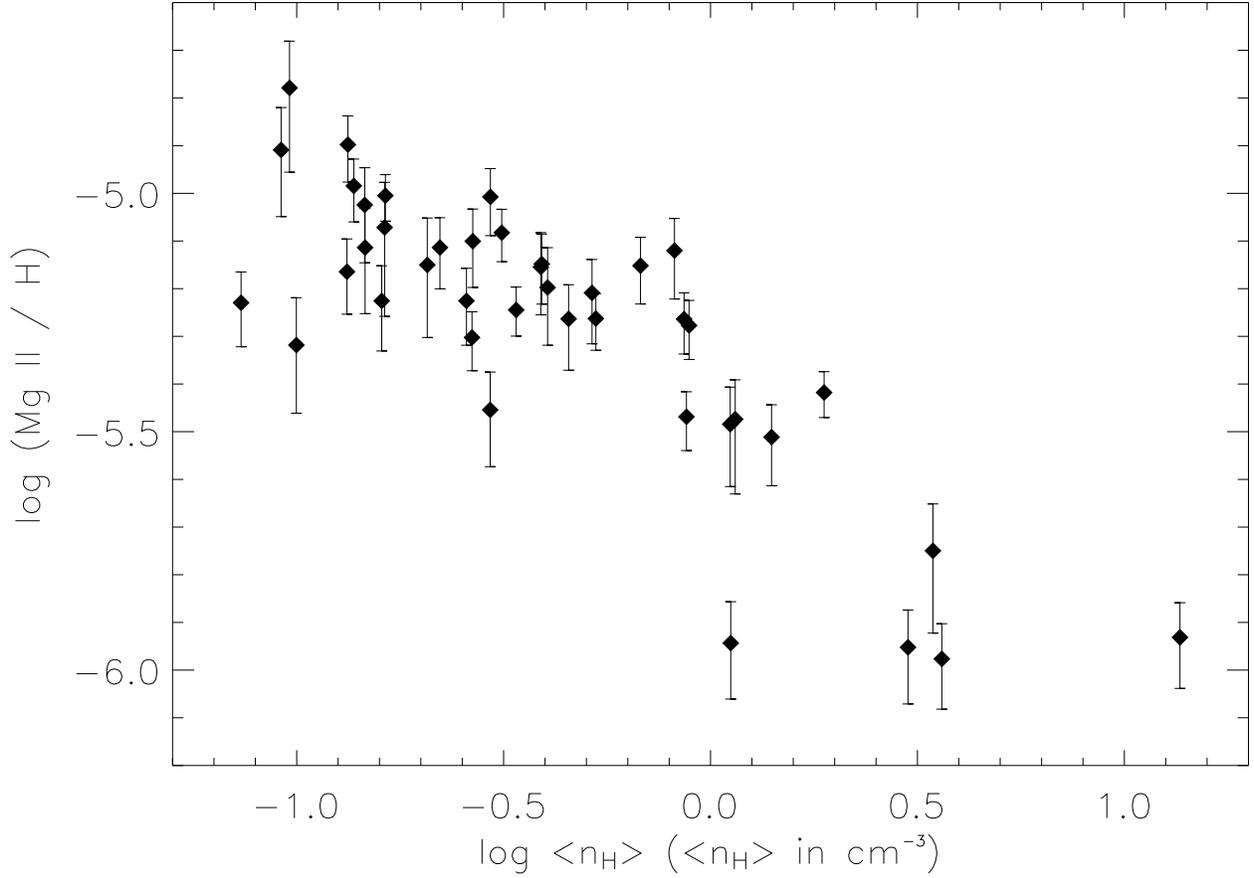}
\end{center}
\caption[$\logMgIIH$ vs. $\nHavg$]{The logarithmic abundance of Mg II relative to hydrogen plotted against the logarithm of the total volume density of hydrogen.  The five lines of sight with the greatest magnesium depletion are (in order of decreasing depletion) HD 27778, HD 37061, HD 37903, HD 147888, and HD 37021.  The Mg II / $\Htot$ ratio and $\nHavg$ for HD 37021 and HD 37061 are based only on atomic hydrogen, due to a lack of data on $\Hmol$.}
\label{fig:logMgIIHlognh}
\end{figure}

First, we note that magnesium depletion is clearly correlated with total hydrogen volume density $\nHavg$.  Five lines of sight (HD 27778, HD 37021, HD 37061, HD 37903, and HD 147888) have substantially larger depletions than any of the other lines of sight in this sample; HD 27778, HD 37021, HD 37061, and HD 147888 are the four densest lines of sight (in terms of $\nHavg$), while the density of HD 37903 is in the top 20\% of our sample.  This correlation is plotted in Figure \ref{fig:logMgIIHlognh}.  Since these dense lines of sight are also some of the lines of sight that are found in both our sample and that of CLMS, we are not probing significantly higher average density and cannot expand on their conclusions regarding $\nHavg$.

We also find that magnesium depletion is correlated with total hydrogen column density.  However, we conclude that this is primarily a secondary correlation due to the correlation with $\nHavg$.  However, we can attempt to analyze whether or not the correlation is independently significant.  To do this, we follow the methods described in \citet{Jenkins1986}.  First, we calculate Pearson correlation coefficients between depletion and the two variables of interest (in this case $\logHtot$ and $\lognHavg$), as well as those two variables with each other.  The partial correlation coefficient is then given by $\rho_{12.3} = (\rho_{12}-\rho_{13}\rho_{23})/[(1-\rho_{13}^2)(1-\rho_{23}^2)]^{-1/2}$, where the subscripts on the correlation coefficients indicate the two variables being correlated, and $\rho_{12.3}$ is the correlation coefficient between the first two variables if the third variable is held fixed.  However, what is really being calculated is $r$, the sample correlation coefficient(s), as opposed to $\rho$, the population correlation coefficient(s).  Once $r_{12.3}$ has been calculated, we examine the significance level for a $t$-test of the appropriate number of degrees of freedom (in this case, the number of data points minus three) to determine the probability that $\rho$ is non-zero (i.e.~a true correlation exists).  Using these methods, we find a 49\% chance of the null hypothesis (i.e.~the two-sided probability that $\rho=0$) for a correlation between Mg II depletion and $\NHtot$ when $\nHavg$ is held fixed.

Because this technique of trivariate analysis uses Pearson correlation coefficients, a few caveats apply, namely that linear relationships and normal distributions are implicitly assumed.  We used $\logMgIIH$, $\logHtot$, and $\lognHavg$ in the analysis just discussed because the correlation between all combinations of those variables are stronger than when in linear form.

CLMS also explored possible correlations between magnesium depletion and $\fHmol$, but only briefly commented on the results.  They concluded that any correlation was less significant than the correlation between magnesium depletion and $\nHavg$, in that it did not as effectively discriminate between the distinct depletion levels expected in the \citet{Spitzer1985} model.  The upper right panel of Figure 9 in CLMS shows a very clear correlation between magnesium depletion and $\fHmol$ for $\fHmol \gtrsim 0.1$, superimposed with a scatter plot in depletion for a smaller subset of lines with $0.01 \lesssim \fHmol \lesssim 0.1$.  Two additional lines of sight with $\fHmol \lesssim 10^{-4}$ conform to the main correlation in that they show minimal depletion.

\begin{figure}[t!]
\begin{center}
\epsscale{1.00}
\plotone{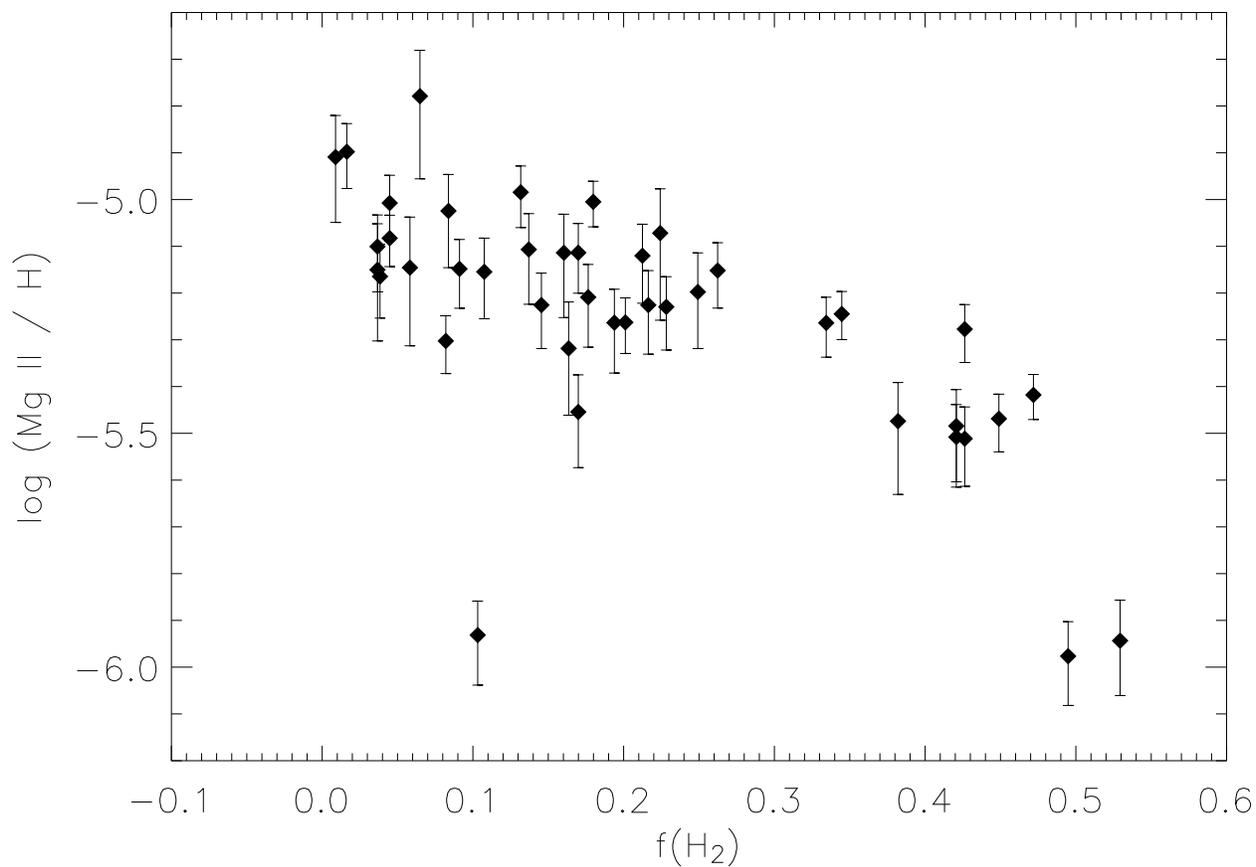}
\end{center}
\caption[$\logMgIIH$ vs. $\nHavg$]{The logarithmic abundance of Mg II relative to hydrogen plotted against the molecular fraction of hydrogen.  Two lines of sight with large depletions are not plotted due to a lack of information on $\Hmol$, HD 37021 and HD 37061.  The anomalous point, with a large depletion but small $\fHmol$, is the HD 147888 line of sight.  Compare this figure to the upper right panel of Figure 9 of CLMS.  See \S \ref{sss:hydrogen_correlations} in this paper for further discussion.}
\label{fig:logMgIIHHf}
\end{figure}

We also see a clear correlation between magnesium depletion and the molecular fraction of hydrogen, $\fHmol$, plotted in Figure \ref{fig:logMgIIHHf}, with depletion increasing with increasing $\fHmol$.  There is the exception of one discrepant point, HD 147888, that exhibits substantial depletion at $\fHmol \sim 0.1$.  We note that CLMS found a larger column density for this line of sight than we do ($0.2\dex$); however, even if we adopt the CLMS Mg II column density for this line of sight, its abundance is still $0.26\dex$ smaller than any line of sight in our sample with $\fHmol < 0.4$.

We concur with CLMS that the correlation between magnesium depletion and $\fHmol$ is not as rigorous as the correlation with $\nHavg$.  We examine the data using various combinations of the variables in their logarithmic (where correlations are the strongest) and linear forms.  In all cases, we find that the probability of the null hypothesis (i.e.~$\rho=0$, as discussed above) between Mg II depletion and $\fHmol$ with $\nHavg$ held constant is less than 5\%; in most cases, it is $\lesssim1$\%.  Conversely, the probability that there is no correlation between Mg II depletion and $\nHavg$ with $\fHmol$ held constant is less than 0.1\%.  Therefore, we conclude that while $\nHavg$ is clearly the dominant correlation, the correlation with $\fHmol$ has some independent significance.

A question of interest is where scatter seems to be introduced and why.  Between Figure 9 of CLMS and Figure \ref{fig:logMgIIHHf} of this paper, scatter is only observed at $\fHmol \lesssim 0.1$.  In our recent related work on Fe II \citep{JensenFeII}, we note that Fe II depletion is also correlated with $\fHmol$, but there are a few exceptions to the trend at both high and low values of $\fHmol$.  The most severe exceptions noted in that paper were HD 147888 and HD 164740 (with large depletions but $\fHmol \lesssim 0.1$) and HD 210121 (with less depletion despite $\fHmol \sim 0.7$).  For Mg II, however, we do not see any outlying points at values of $\fHmol$ larger than $\sim0.1$ within either this sample or CLMS.  Points such as HD 147888, however, still require explanation.

\citet{Snow1983} put forth the possibility that in some dense environments an increased average grain size, which decreases the grain surface area per unit volume, may suppress $\Hmol$ formation (as $\Hmol$ is thought to form on grain surfaces).  This scenario was specifically discussed in the context of the $\rho$ Oph cloud, for which there is indepedent evidence (in part, a value of $\rv$ greater than the interstellar average of 3.1) that grain coagulation has occurred.  Our main outlying line of sight, HD 147888 (with $\rv$ of 4.06), passes through the $\rho$ Oph cloud, so observing the combination of a dense, depleted environment with small $\fHmol$ is not surprising in this case.

Two of the other lines of sight with large depletions are HD 37021 and HD 37061.  Reliable far-ultraviolet data sets do not exist for these lines of sight; therefore, they do not have measurements of the molecular hydrogen column densities or subsequently derived molecular fractions of hydrogen.  However, as stated above in \S \ref{ss:hydrogen}, \citet{Cartledge2001} has argued that these lines of sight have small values of $\fHmol$ based on a lack of Cl I.  These two lines of sight also have large values of $\rv$ (5.54 and 4.23, respectively), implying a larger average grain size.  The possible effect of grain size on depletions and $\Hmol$ formation, and other interpretations, will be discussed further in \S \ref{sss:extinction_correlations}.

Barring such outlying points, the trend of increased Mg II depletion with increased $\fHmol$ has a relatively clear interpretation.  $\Hmol$ is formed in the same dense, dusty environments that foster large depletions.  Within the context of this sample and CLMS, this seems to hold for lines of sight with $\fHmol \gtrsim 0.1$.  While the ubiquity of $\Hmol$ even in diffuse regions complicates the issue \citep[see conclusions of][]{Rachford2002}, there is still a physical argument that $\fHmol$ should be a good diagnostic of the local conditions of interstellar clouds, and therefore depletions.  It is worth noting that the similar trend between iron depletion and $\fHmol$ exhibits scatter up to $\fHmol \sim 0.3$ in the work of \citet{SavageBohlin}, in addition to the outlying points mentioned above from \citet{JensenFeII}.  Whether the range in $\fHmol$ over which there is scatter in the abundances is truly different for magnesium and iron or is simply a selection effect is unclear.


\subsubsection{Correlations with Extinction and Reddening Parameters}
\label{sss:extinction_correlations}
\begin{figure}[t!]
\begin{center}
\epsscale{1.00}
\plotone{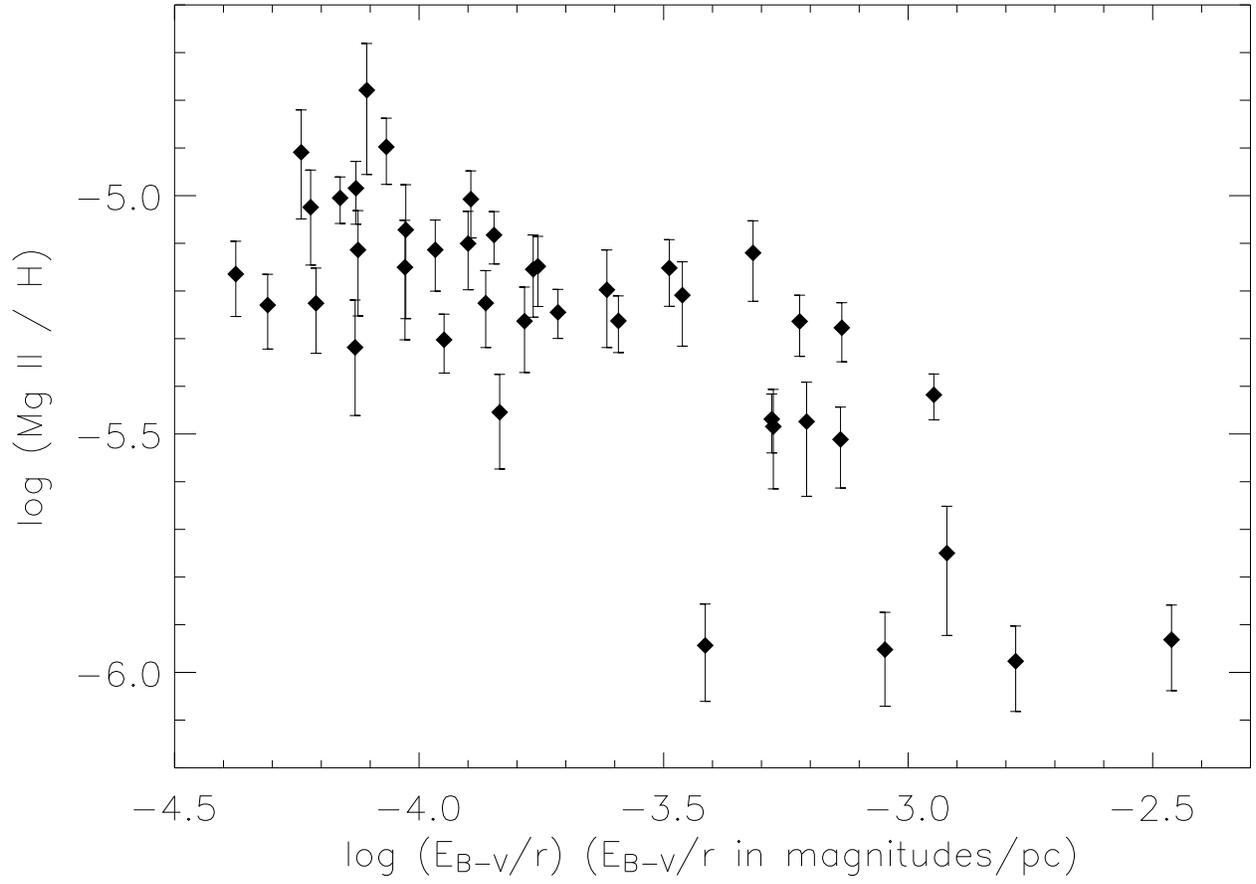}
\end{center}
\caption[$\logMgIIH$ vs. $\log{\ebvdist}$]{The logarithmic abundance of Mg II relative to hydrogen plotted against the logarithm of selective extinction divided by pathlength.  See \S \ref{sss:extinction_correlations} for further discussion.}
\label{fig:logMgIIHebv_dist}
\end{figure}

\begin{figure}[t!]
\begin{center}
\epsscale{1.00}
\plotone{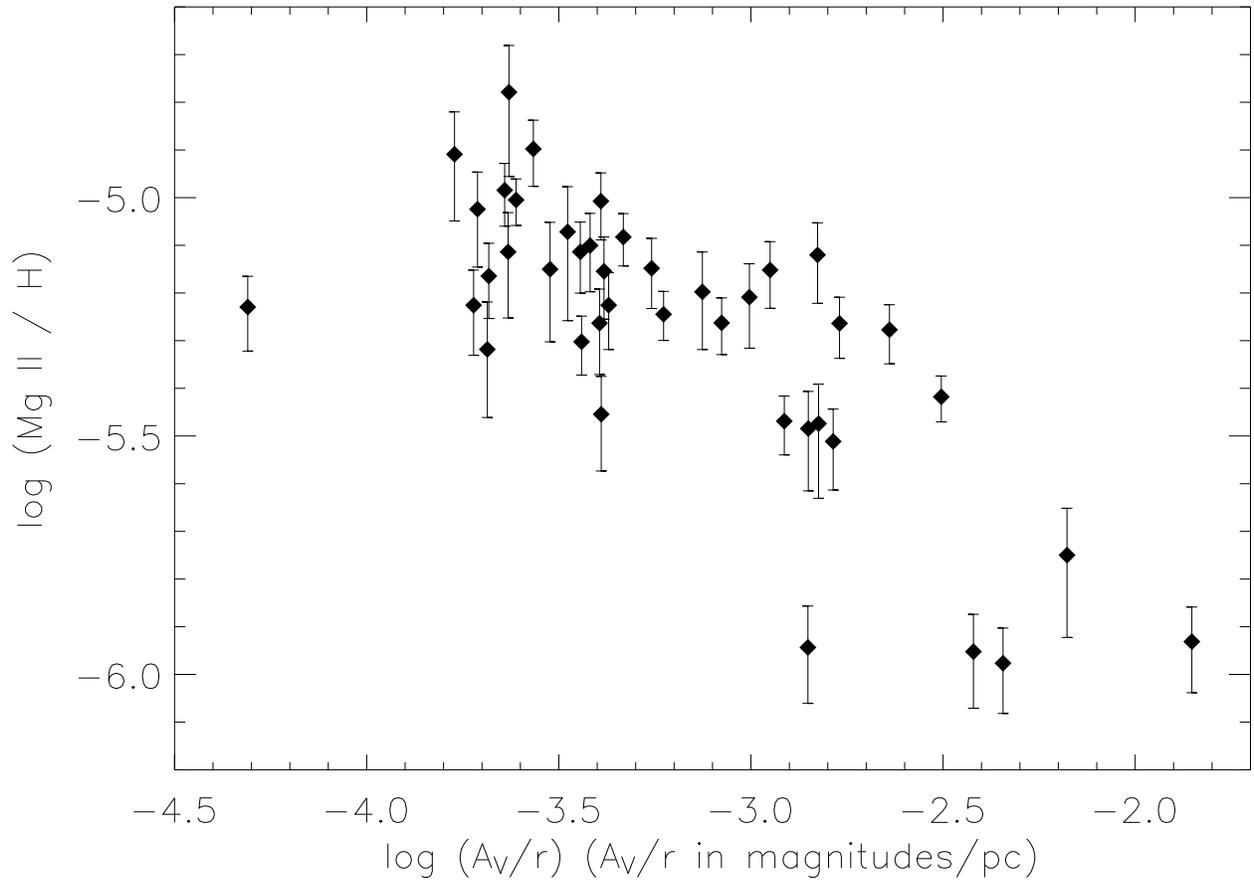}
\end{center}
\caption[$\logMgIIH$ vs. $\log{\avdist}$]{The logarithmic abundance of Mg II relative to hydrogen plotted against the logarithm of total visual extinction divided by pathlength.  See \S \ref{sss:extinction_correlations} for further discussion.}
\label{fig:logMgIIHav_dist}
\end{figure}

We find that the depletion of Mg II is correlated to both selective extinction, $\ebv$, and total visual extinction, $\av$.  However, there is significant scatter in these correlations.  Because these are integrated line of sight parameters, it makes sense to divide by line-of-sight pathlength.  Both $\ebv$ and $\av$ are strongly correlated with $\NHtot$, and both are thought to be rough measures of the total dust column density.  Therefore, $\ebvdist$ and $\avdist$ should be strongly correlated with $\nHavg$ and be approximations of the total dust volume density.  When we look for correlations between magnesium depletion and $\ebvdist$ and $\avdist$ we find that the correlations are substantially increased when compared to the integrated line of sight parameters.  Therefore, we can say, with reasonable confidence, that magnesium depletion is increased in increasingly dusty environments.  The correlations with $\ebvdist$ and $\avdist$ are plotted in Figures \ref{fig:logMgIIHebv_dist} and \ref{fig:logMgIIHav_dist}.

As with the correlation between depletion and $\fHmol$, we examine partial correlation coefficients to determine the independent significance of these correlations.  The partial correlation coefficient between Mg II depletion and $\log{\ebvdist}$ with $\lognHavg$ held fixed implies that the probability of the null hypothesis is less than 6\%.  The same partial correlation coefficient with $\log{\avdist}$ implies that there is less than a 1\% chance of the null hypothesis.  (Note that both probabilities are two-sided to a $t$-distribution.)  We consider these variables in their logarithmic forms because these are the versions of the variables that exhibit the strong correlations (for all combinations of the variables in question).  However, if the variables are considered in non-logarithmic forms, the probability of the null hypothesis generally increases.  Again we note that this type of trivariate statistical measure implicitly assumes that the correlation is linear with normally distributed scatter.  As with the correlation between depletion and $\fHmol$, we conclude that while these correlations do not improve on $\nHavg$ as a predictor of depletions, this is limited evidence that they are significant in their own right.


We have briefly explored the possibility that the additional (or ``missing'') magnesium that is depleted from these lines of sight is found in the form of gas-phase Mg I in regions that are presumably shielded from radiation by dust.  While the STIS data do not cover Mg I absorption lines in many cases, our results indicate that gas-phase Mg I column densities are far too small to account for the order-of-magnitude increase in depletion seen in gas-phase Mg II.  Therefore, it likely that the missing gas-phase magnesium is tied up in the additional grains found in these environments, supporting the conclusion above that the correlations between Mg II depletion and the parameters $\log{\ebvdist}$ and $\log{\avdist}$ are physically significant.  It is also worth noting that in no case do we see Mg/H less than $\sim1\ppm$, even in the densest environments, with values of $\ebvdist$ and $\avdist$ several times larger than the average of the sample.  This suggests that these lines of sight are not probing what might be considered ``translucent clouds'' \citep[though some may be ``translucent lines of sight''; see][]{SnowMcCall}.

\begin{figure}[t!]
\begin{center}
\epsscale{1.00}
\plotone{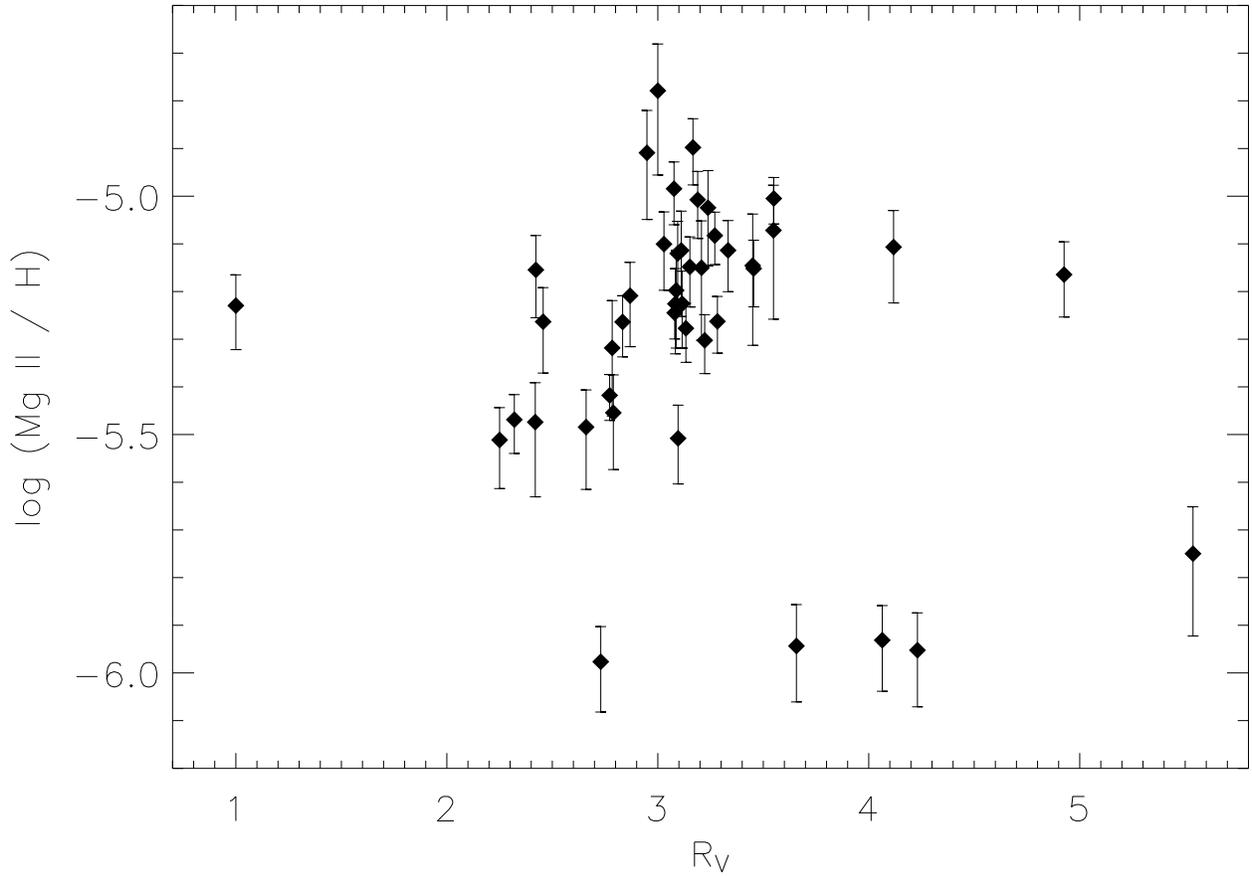}
\end{center}
\caption[$\logMgIIH$ vs. $\rv$]{The logarithmic abundance of Mg II relative to hydrogen plotted against the ratio of total visual extinction to selective extinction.  Trends are visually less apparent than for other parameters, but a reasonably strong linear correlation exists between magnesium depletion and $\rv$, which implies that magnesium depletion is correlated with increasing grain size.  See \S \ref{sss:extinction_correlations} for further discussion.}
\label{fig:logMgIIHRv}
\end{figure}

The Mg II abundance is plotted against the ratio of total visual to selective extinction, $\rv \equiv \av / \ebv$, in Figure \ref{fig:logMgIIHRv}.  The plot shows significant scatter; however, some statistical measures show the possibility of a slight correlation.  A Pearson correlation coefficient between $\logMgIIH$ and $\rv$ implies that the probability of the null hypothesis is about 31\%.  A Spearman's $\rho$ rank correlation coefficient, which does not depend on the functional form assumed (including whether variables are considered linearly or logarithmically) beyond assuming that the correlation is either monotonically increasing or monotonically decreasing, shows a negative correlation (decreasing abundance/increasing depletion as $\rv$ increases) and is significant to approximately 1.3-$\sigma$.  Therefore, there is evidence of a possible slight correlation between depletion and $\rv$.  However, this is far from a certain conclusion.  Significant selection effects are also a possibility, as the correlations are dominated by some of the points that have greater depletion and large values of $\rv$.  In fact, if these lines of sight are excluded, the trend begins to reverse toward a positive correlation between increasing Mg II abundance and increasing $\rv$.  In general, we conclude that $\rv$ is a poor predictor of depletions; as one anecdotal counterexample, HD 91597 has a very large value of $\rv=4.9$ but does not exhibit particularly large Mg II depletion.

However, there are a few lines of sight that present interesting interpretive challenges where the value of $\rv$ may provide insight.  As discussed above, the possibility of large magnesium depletion but small $\fHmol$ exists for three lines of sight:  HD 37021, HD 37061, and HD 147888.  In the latter case the effect is clear, while in the former two cases the small value of $\fHmol$ is merely inferred.  What is interesting, as noted above, is that these three lines of sight all have values of $\rv>4$ which is a fairly significant deviation from the interstellar average of 3.1.  Because large grains contribute to $\av$ (i.e.~grey extinction) but less so to $\ebv$, $\rv$ is thought to be correlated to average grain size.  Explaining why depletion should increase in a line of sight with large grains is difficult.  As mentioned above in our discussion of $\Hmol$ and iron depletion, increased grain size decreases dust surface area per unit volume, and therefore reduces $\Hmol$ formation rates.  However, decreased surface area per unit volume also implies a reduction in rates of sticking between dust grains and gas-phase atoms.

It seems we can reasonably conclude that the large values of $\rv$, i.e.~the larger average grain populations, are not responsible for the large depletions by way of atoms and ions sticking to the grains.  One possibility is that the large depletions are instead ``locked in'' prior to grain coagulation.  Another possibility is the effect of a high-radiation field:  this is known for the line of sight toward HD 147888 ($\rho$ Oph D) as well as HD 37021 and HD 37061 which are in Orion (radiation is presumed to be responsible for the relative lack of Cl I, and thus also $\Hmol$, as mentioned in \S \ref{ss:hydrogen}).  However, \citet{Snow1983} argues that the increased radiation is unlikely to be entirely responsible for the low $\fHmol$ in the $\rho$ Oph cloud.  Whether or not this is the case for HD 37021 and HD 37061 is unclear.  More details about the radiation field and the exact nature of the grain population (we have only considered the crude measure of $\rv$) are probably necessary to fully understand these lines of sight.




\subsubsection{Anticorrelation with Distance}
\label{sss:distance_anticorrelation}
\begin{figure}[t!]
\begin{center}
\epsscale{1.00}
\plotone{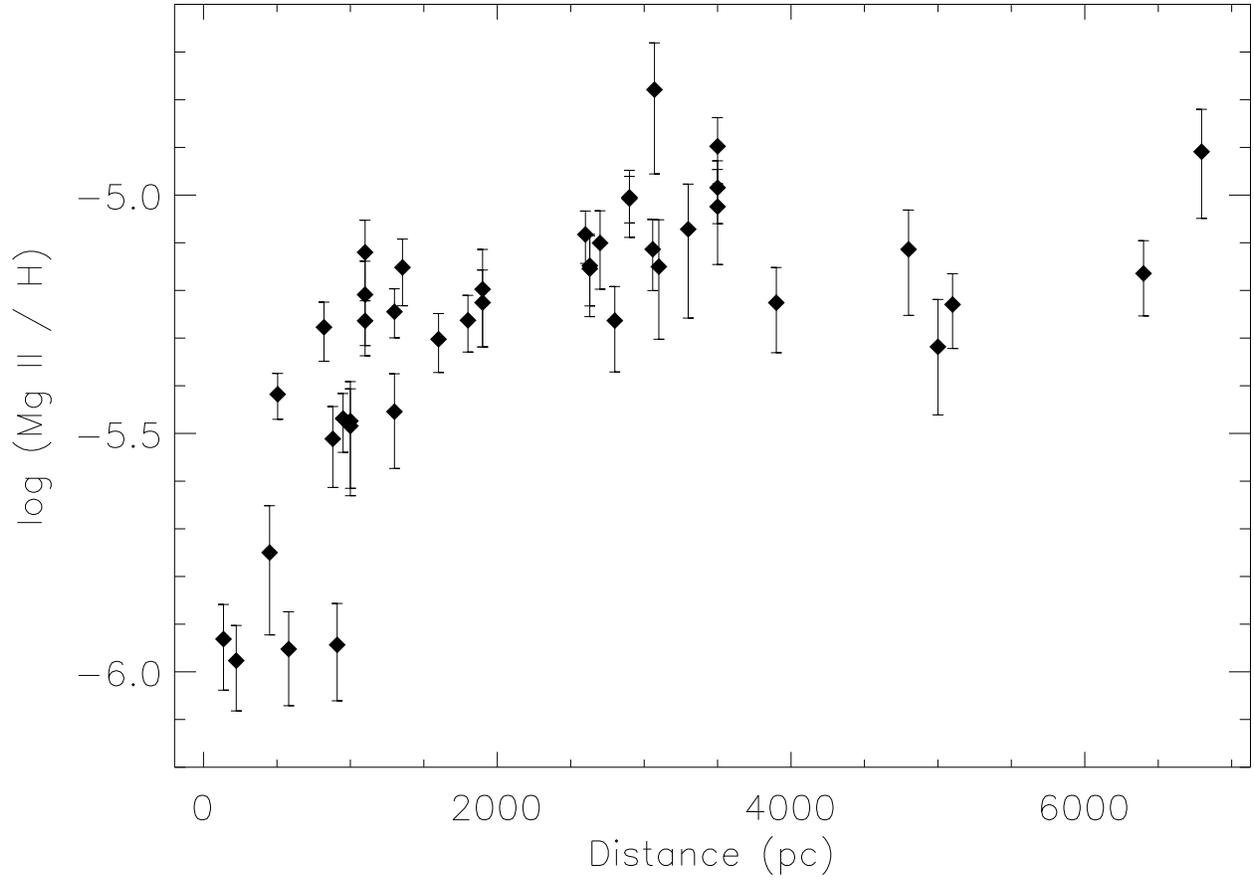}
\end{center}
\caption[$\logMgIIH$ vs. Distance]{The logarithmic abundance of Mg II relative to hydrogen plotted against pathlength.  Note the trend of decreasing depletion up to about 1.5--2 kpc, and relatively constant depletion beyond 2 kpc.}
\label{fig:logMgIIHdist}
\end{figure}

We find that magnesium depletion is generally anticorrelated with distance to the background star, that is, line of sight pathlength; this relationship is shown in Figure \ref{fig:logMgIIHdist}.  Depletion decreases by nearly an order of magnitude between very short lines of sight and those up to about 2 kpc or so, and then is relatively constant (to within about 0.3-0.4$\dex$) out to about 6 kpc.  As we concluded for a similar anticorrelation seen between iron depletion and distance \citep{JensenFeII}, the long pathlengths are likely sampling a variety of cloud conditions, resulting in the constant depletion for long-pathlength lines of sight.  On the other hand, given the comparable hydrogen column densities of all the lines of sight in this study ($\logHtot \approx21-22$), the shorter lines of sight are generally the denser lines of sight.

\subsubsection{Spatial Variations}
\label{sss:spatial_variations}
We find one very interesting correlation with Galactic location:  the five stars with the largest depletions reside at higher Galactic latitudes of $|b|>15^{\circ}$.  However, with pathlengths of less than 1 kpc, these lines of sight are still primarily in the Galactic disk.  When we analyze magnesium depletions against the height from the center of the Galactic disk, $z=r \sin{b}$, we do not see a strong correlation.  The variation with respect to Galactic latitude is most likely a coincidence, given that these are some of the densest and most reddened lines of sight.  We do not see any other evidence of significant spatial variations.


\subsection{Mg/H of Galactic Stars}
\label{ss:stellarMgH}
In the last several years, three major papers have attempted to analyze the cosmic abundance ``standards'' in the ISM through studies of stellar abundances and meteoritic abundances---\citet{SnowWitt}, \citet{SofiaMeyer}, and \citet{Lodders}.  The importance of these standards is to compare them with the observed gas-phase abundances and infer an absolute value for depletions---and therefore absolute values for the amount of these elements in phases other than atomic gas, i.e.~dust grains and molecules.

Of the major elements relevant to dust, the element with the best determined cosmic abundance is iron.  The four major measurements of the cosmic Fe/H ratio---solar, B stars, F and G stars, and CI chondrites---all agree very closely, largely within the errors.  The situation is somewhat more complex for other elements.  Carbon and oxygen show apparent overabundances in the Sun compared to F and G stars (whether or not the solar and F/G star abundances potentially agree within the errors depends on the choice of solar abundances, regarding which there is still some uncertainty), while B stars show relative deficits in these abundances compared to the Sun and other F and G stars.  The chondritic abundances of C and O are even smaller.  Nitrogen seems to be somewhat less abundant in B stars than in the Sun, though the two are reconciliable within the errors; F and G nitrogen abundances are generally unknown, and the chondritic abundances are substantially lower.  Silicon seems to be most abundant in F and G stars, slightly less abundant in the Sun, and about half as abundant in B stars.  The errors, however, do not rule out agreement between all three measurements.  However, the chondritic abundance tightly matches the solar abundance.

Both \citet{SofiaMeyer} and \citet{Lodders}, cite \citet{Holweger} for the solar abundance of magnesium, $\logMgH=-4.46$; \citet{SnowWitt} report a slightly older value from \citet{AndersGrevesse} of $\logMgH=-4.42$, though these values are consistent within the errors.  The chondritic abundances in \citet{Lodders} of $\logMgH=-4.44$ are also very consistent with these solar values.  The discrepancy arises when various stellar abundances are considered.  Both \citet{SnowWitt} and \citet{SofiaMeyer} found significantly smaller abundances of Mg for B stars.  \citet{SnowWitt} found $\logMgH=-4.63$ for field B stars and $\logMgH=-4.68$ for cluster B stars; \citet{SofiaMeyer}, making no distinction between cluster and field stars, found $\logMgH=-4.64$ for all B stars.  Though there is marginal agreement within the very large errors in these numbers, the B star abundances are $\approx60\%$ smaller than the solar abundances.  \citet{SnowWitt} and \citet{SofiaMeyer} also  disagree on the Mg abundance in F and G stars ($\logMgH=-4.52$ and $-4.37$, respectively) due most likely to \citet{SofiaMeyer} restricting their sample to stars with ages of $\leq2$ Gyr.  Again, however, these values have relatively large errors and are reconciliable with the B star abundances, though just barely.

What effect does the choice of a cosmic magnesium abundance have for the implied dust-phase abundances to be used in dust models?  The differences between the cosmic abundances just discussed leads to nontrivial differences in the dust-phase abundances.  Our weighted interstellar average of Mg II/H is $2.7\pm0.1\ppm$ (parts per million), though the median value in our 44 lines of sight is somewhat larger at $6.2\ppm$.  Taking the extremes of the above numbers, anywhere between $\sim20$ and $\sim40\ppm$ of Mg is available for creating dust.  Examining the various models compared in Table 3 of \citet{SnowWitt}, we find that most models require much more Mg than the lower value of $\sim20$ implied by a B star abundance standard.  That a B star abundance is less likely to represent the cosmic abundance was also found by \citet{ZDA2004}, who had more difficulty fitting dust models to observations using the dust-phase abundances implied by assuming B star abundances as the cosmic standard.  In fact, this is true even though \citeauthor{ZDA2004} assumed $\approx0$ ppm of magnesium to be in the gas-phase.  If the few ppm of magnesium in the gas-phase as measured by this paper and CLMS were included, the \citet{ZDA2004} fits would become even more strained (the best fits for B star abundances were at the limit of the error in those abundances and inferior to the fits obtained using other abundances).  Therefore, the major conclusion that we can make regarding cosmic abundances and the incorporation of magnesium into dust is to add to the evidence that B star abundances, despite B stars being younger and therefore potentially good tracers of the current ISM, are a poor cosmic standard.  Whether or not the solar or an F and G star abundance standard for Mg is a better fit is a test that is too sensitive for us to comment on, given the uncertainties in those abundances.


\section{SUMMARY}
\label{s:MgII_summary}
We have analyzed the abundance of Mg II in 44 lines of sight.  Our study does not probe substantially larger average hydrogen volume densities than previously observed by CLMS; therefore, we observe the same correlations between Mg II and the $\nHavg$ and $\ebvdist$.  We also note a correlations between magnesium depletion and $\avdist$, a different measure of dust density.  Correlations between $\NHtot$ and the reddening and extinction parameters $\ebv$ and $\av$ mean that correlations between Mg II depletion and dust density measures are expected.  However, these latter correlations, while not strong than the correlation between depletion and $\nHavg$, show some evidence of being significant even at fixed $\nHavg$ and should be more directly related to the line-of-sight dust content.  We also note a correlation between magnesium depletion and $\fHmol$ in our data; combined with the results of CLMS, this correlation seems to be valid for $\fHmol \gtrsim 0.1$ but not at smaller $\fHmol$.  A question that is related to the trend with $\fHmol$ is why so little $\Hmol$ forms in certain high-density lines of sight.  Our results are consistent with the \citet{Snow1983} suggestion that the reduced grain surface area per unit volume of large grains plays a role in reducing $\Hmol$ formation rates.  For similar reasons, we can conclude that the grain coagulation probably occurs after depletions are already ``locked into'' the dust, rather than depletion of gas-phase atoms onto grain surfaces.

\acknowledgements

The authors would like to thank the anonymous referee for many helpful comments that greatly improved the manuscript.  The authors also wish to thank S.~V.~Penton for the use of his measurements of the STIS point-spread functions; B.~A.~Keeney for the original version of the profile-fitting code that we adapted to our purposes; and J.~M.~Shull and B.~L.~Rachford for the use of their unpublished measurements of molecular hydrogen column densities.  This work was supported by NASA grant NAG5-12279.

\bibliographystyle{apj}
\bibliography{refs}

\clearpage \clearpage

\begin{deluxetable}{cccccc}
\tablecolumns{6}
\tablewidth{0pc}
\tabletypesize{\tiny}
\tablecaption{Lines of Sight: Stellar Data\label{MgII_stellardata}}
\tablehead{\colhead{Star Name} & \colhead{Spectral Class} & \colhead{l} & \colhead{b} & \colhead{Distance (pc)} & \colhead{Ref.}}
\startdata
BD +35$^{\circ}$4258 & B0.5Vn & 77.19 & -4.74 & 3100 & 1 \\
CPD -59$^{\circ}$2603 & O7V... & 287.59 & -0.69 & 2630 & 2 \\
HD 12323 & O9V & 132.91 & -5.87 & 3900 & 1 \\
HD 13745 & O9.7II((N)) & 134.58 & -4.96 & 1900 & 1 \\
HD 15137 & O9.5V & 137.46 & +7.58 & 3300 & 2 \\
HD 27778 & B3V & 172.76 & -17.39 & 223 & 3 \\
HD 37021 & B0V & 209.01 & -19.38 & 450 & 4 \\
HD 37061 & B1V & 208.92 & -19.27 & 580 & 5 \\
HD 37903 & B1.5V & 206.85 & -16.54 & 910 & 5 \\
HD 40893 & B0IV & 180.09 & +4.34 & 2800 & 5 \\
HD 66788 & O8/O9Ib & 245.43 & +2.05 & \nodata & \nodata \\
HD 69106 & B1/B2II & 254.52 & -1.33 & 1600 & 5 \\
HD 91597 & B7/B8IV/V & 286.86 & -2.37 & 6400 & 5 \\
HD 91651 & O9VP: & 286.55 & -1.72 & 3500 & 1 \\
HD 92554 & O9.5III & 287.60 & -2.02 & 6795 & 2 \\
HD 93205 & O3V & 287.57 & -0.71 & 2600 & 1 \\
HD 93222 & O7III((f)) & 287.74 & -1.02 & 2900 & 1 \\
HD 93843 & O6III(f) & 228.24 & -0.90 & 2700 & 1 \\
HD 94493 & B0.5Ib & 289.01 & -1.18 & 2900 & 5 \\
HD 99857 & B1Ib & 294.78 & -4.94 & 3058 & 2 \\
HD 99890 & B0.5V: & 291.75 & +4.43 & 3070 & 2 \\
HD 103779 & B0.5II & 296.85 & -1.02 & 3500 & 5 \\
HD 104705 & B0.5III & 297.45 & -0.34 & 3500 & 5 \\
HD 109399 & B1Ib & 301.71 & -9.88 & 1900 & 5 \\
HD 116781 & O9/B1(I)E & 307.05 & -0.07 & \nodata & \nodata \\
HD 122879 & B0Ia & 312.26 & +1.79 & 4800 & 5 \\
HD 124314 & O7 & 312.67 & -0.42 & 1100 & 1 \\
HD 147888 & B3/B4V & 353.65 & +17.71 & 136 & 3 \\
HD 152590 & O7.5V & 344.84 & +1.83 & 1800 & 4 \\
HD 168941 & B0III/IV & 5.82 & -6.31 & 5000 & 5 \\
HD 177989 & B2II & 17.81 & -11.88 & 5100 & 5 \\
HD 185418 & B0.5V & 53.60 & -2.17 & 950 & 5 \\
HD 192639 & O7.5IIIF & 74.90 & +1.48 & 1100 & 5 \\
HD 195965 & B0V & 85.71 & +5.00 & 1300 & 5 \\
HD 202347 & B1V & 88.22 & -2.08 & 1300 & 1 \\
HD 203374 & B0IVpe & 100.51 & +8.62 & 820 & 2 \\
HD 206267 & O6(F) & 99.29 & +3.74 & 1000 & 4 \\
HD 207198 & O9II & 103.14 & +6.99 & 1000 & 5 \\
HD 207308 & B0.5V & 103.11 & +6.82 & \nodata & \nodata \\
HD 207538 & O9.5V & 101.60 & +4.67 & 880 & 5 \\
HD 209339 & BOIV & 104.58 & +5.87 & 1100 & 1 \\
HD 210839 & O6e & 103.83 & +2.61 & 505 & 3 \\
HD 224151 & B0.5IISBV & 115.44 & -4.64 & 1355 & 2 \\
HD 303308 & O3V & 287.59 & -0.61 & 2630 & 2 \\
\enddata
\tablerefs{Spectral classes and Galactic coordinates compiled from SIMBAD database at http://simbad.u-strasbg.fr/simbad/sim-fid.  References for distances:  (1) \citet{Savage1985}.  (2) \citet{DS1994}.  (3) Hipparcos 4-$\sigma$---Hipparcos parallax of 4-$\sigma$ precision or better.  (4) Member of an OB association, cluster, or multiple-star system, DIB database at http://dib.uiuc.edu; values compiled from the literature or derived by L.~M.~Hobbs.  (5) Spectroscopic distance modulus from the DIB database.}
\end{deluxetable}

\clearpage \clearpage

\begin{deluxetable}{cccc}
\tablecolumns{4}
\tablewidth{0pc}
\tabletypesize{\scriptsize}
\tablecaption{Lines of Sight:  STIS Data Sets Used\label{MgII_obstable}}
\tablehead{\colhead{Star Name} & \colhead{Data Set(s)} & \colhead{Grating} & \colhead{Aperture\tablenotemark{a}}}
\startdata
BD +35$^{\circ}$4258 & O6LZ89010 & E140M & 0.2X0.2 \\
CPD -59$^{\circ}$2603 & O4QX03010 & E140H & 0.2X0.09 \\
HD 12323 & O63505010 & E140M & 0.2X0.2 \\
HD 13745 & O6LZ05010 & E140M & 0.2X0.2 \\
HD 15137 & O5LH02010 & E140H & 0.1X0.03 \\
 & O5LH02020 & & \\
 & O5LH02030 & & \\
 & O5LH02040 & & \\
HD 27778 & O59S01010 & E140H & 0.2X0.09 \\
 & O59S01020 & & \\
HD 37021 & O59S02010 & E140H & 0.2X0.09 \\
HD 37061 & O59S03010 & E140H & 0.1X0.03 \\
HD 37903 & O59S04010 & E140H & 0.2X0.09 \\
HD 40893 & O8NA02010 & E140H & 0.2X0.2 \\
 & O8NA02020 &  &  \\
HD 66788 & O6LZ26010 & E140M & 0.2X0.2 \\
HD 69106 & O5LH03010 & E140H & 0.1X0.03 \\
 & O5LH03020 & &  \\
HD 91597 & O6LZ32010 & E140M & 0.2X0.2 \\
HD 91651 & O6LZ34010 & E140M & 0.2X0.2 \\
HD 92554 & O6LZ36010 & E140M & 0.2X0.2 \\
HD 93205 & O4QX01010 & E140H & 0.2X0.09 \\
HD 93222 & O4QX02010 & E140H & 0.2X0.09 \\
HD 93843 & O5LH04010 & E140H & 0.1X0.03 \\
 & O5LH04020 & &  \\
HD 94493 & O54306010 & E140H & 0.1X0.03 \\
HD 99857 & O54301010 & E140H & 0.1X0.03 \\
 & O54301020 & & \\
 & O54301030 & & \\
HD 99890 & O6LZ45010 & E140M & 0.2X0.2 \\
HD 103779 & O54302010 & E140H & 0.1X0.03 \\
HD 104705 & O57R01010 & E140H & 0.2X0.09 \\
HD 109399 & O54303010 & E140H & 0.1X0.03 \\
HD 116781 & O5LH05010 & E140H & 0.1X0.3 \\
 & O5LH05020 & & \\
HD 122879 & O5LH07010 & E140H & 0.1X0.03 \\
 & O5LH07020 & & \\
HD 124314 & O54307010 & E140H & 0.1X0.03 \\
HD 147888 & O59S05010 & E140H & 0.2X0.09 \\
HD 152590 & O8NA04010 & E140H & 0.2X0.2 \\
 & O8NA04020 & & \\
 & O5C08P010 & & \\
HD 168941 & O6LZ81010 & E140M & 0.2X0.2 \\
HD 177989 & O57R03010 & E140H & 0.2X0.09 \\
 & O57R03020 &  & \\
HD 185418 & O5C01Q010 & E140H & 0.2X0.2 \\
HD 192639 & O5C08T010 & E140H & 0.2X0.2 \\
HD 195965 & O6BG01010 & E140H & 0.1X0.03 \\
HD 202347 & O5G301010 & E140H & 0.1X0.03 \\
HD 203374 & O5LH08010 & E140H & 0.1X0.03 \\
 & O5LH08020 & & \\
 & O5LH08030 & & \\
HD 206267 & O5LH09010 & E140H & 0.1X0.03 \\
 & O5LH09020 & & \\
HD 207198 & O59S06010 & E140H & 0.2X0.09 \\
 & O59S06020 & & \\
HD 207308 & O63Y02010 & E140M & 0.2X0.06 \\
 & O63Y02020 & & \\
HD 207538 & O63Y01010 & E140M & 0.2X0.06 \\
 & O63Y01020 & & \\
HD 209339 & O5LH0B010 & E140H & 0.1X0.03 \\
 & O5LH0B020 & & \\
HD 210839 & O54304010 & E140H & 0.1X0.03 \\
HD 224151 & O54308010 & E140H & 0.1X0.03 \\
HD 303308 & O4QX04010 & E140H & 0.2X0.09 \\
\enddata
\tablenotetext{a}{Aperture is length X width of the slit, with both values in arcseconds.}
\end{deluxetable}

\clearpage \clearpage

\begin{deluxetable}{cccccccc}
\tablecolumns{8}
\tablewidth{0pc}
\tabletypesize{\tiny}
\tablecaption{Lines of Sight: Hydrogen Data\label{MgII_hydrogentable}}
\tablehead{\colhead{Line of Sight} & \colhead{$\logHI$} & \colhead{Ref.} & \colhead{$\logHmol$} & \colhead{Ref.} & \colhead{$\logHtot$} & \colhead{$\nHavg$} & \colhead{$\fHmol$}}
\startdata
BD +35$^{\circ}$4258 & $21.28\pm0.10$ & 1 & $19.56\pm0.03$ & 2 & $21.30\pm0.10$ & $0.21$ & $0.04$ \\
CPD -59$^{\circ}$2603 & $21.46\pm0.07$ & 3 & $20.16\pm0.03$ & 2 & $21.50\pm0.06$ & $0.39$ & $0.09$ \\
HD 12323 & $21.18\pm0.09$ & 4 & $20.32\pm0.08$ & 4 & $21.29\pm0.07$ & $0.16$ & $0.22$ \\
HD 13745 & $21.25\pm0.10$ & 3 & $20.47\pm0.05$ & 2 & $21.37^{+0.08}_{-0.07}$ & $0.40$ & $0.25$ \\
HD 15137 & $21.11\pm0.16$ & 3 & $20.27\pm0.03$ & 2 & $21.22^{+0.13}_{-0.12}$ & $0.16$ & $0.22$ \\
HD 27778 & $21.10\pm0.12$ & 4 & $20.79\pm0.06$ & 5 & $21.40^{+0.07}_{-0.06}$ & $3.62$ & $0.49$ \\
HD 37021 & $21.68\pm0.12$ & 4 & \nodata & 0 & $21.68\pm0.12$ & $3.45$ & \nodata \\
HD 37061 & $21.73\pm0.09$ & 4 & \nodata & 0 & $21.73\pm0.09$ & $3.00$ & \nodata \\
HD 37903 & $21.17\pm0.10$ & 3 & $20.92\pm0.06$ & 6 & $21.50^{+0.06}_{-0.05}$ & $1.12$ & $0.53$ \\
HD 40893 & $21.50\pm0.10$ & 7 & $20.58\pm0.05$ & 6 & $21.59\pm0.08$ & $0.45$ & $0.19$ \\
HD 66788 & $21.23\pm0.10$ & 1 & $19.72\pm0.03$ & 2 & $21.26\pm0.09$ & \nodata & $0.06$ \\
HD 69106 & $21.08\pm0.06$ & 3 & $19.73\pm0.05$ & 2 & $21.12^{+0.06}_{-0.05}$ & $0.27$ & $0.08$ \\
HD 91597 & $21.40\pm0.06$ & 3 & $19.70\pm0.05$ & 2 & $21.42\pm0.06$ & $0.13$ & $0.04$ \\
HD 91651 & $21.15\pm0.06$ & 3 & $19.07\pm0.03$ & 2 & $21.16\pm0.06$ & $0.13$ & $0.02$ \\
HD 92554 & $21.28\pm0.10$ & 3 & $18.93\pm0.05$ & 2 & $21.28\pm0.10$ & $0.09$ & $0.01$ \\
HD 93205 & $21.38\pm0.05$ & 8 & $19.75\pm0.03$ & 2 & $21.40\pm0.05$ & $0.31$ & $0.04$ \\
HD 93222 & $21.40\pm0.07$ & 8 & $19.77\pm0.03$ & 2 & $21.42\pm0.07$ & $0.29$ & $0.04$ \\
HD 93843 & $21.33\pm0.08$ & 3 & $19.61\pm0.03$ & 2 & $21.35\pm0.08$ & $0.27$ & $0.04$ \\
HD 94493 & $21.08\pm0.05$ & 8 & $20.12\pm0.05$ & 9 & $21.17\pm0.04$ & $0.16$ & $0.18$ \\
HD 99857 & $21.24\pm0.08$ & 8 & $20.25\pm0.05$ & 2 & $21.32\pm0.07$ & $0.22$ & $0.17$ \\
HD 99890 & $20.93\pm0.13$ & 3 & $19.47\pm0.05$ & 2 & $20.96\pm0.12$ & $0.10$ & $0.06$ \\
HD 103779 & $21.16\pm0.10$ & 3 & $19.82\pm0.05$ & 2 & $21.20\pm0.09$ & $0.15$ & $0.08$ \\
HD 104705 & $21.11\pm0.07$ & 3 & $19.99\pm0.03$ & 2 & $21.17\pm0.06$ & $0.14$ & $0.13$ \\
HD 109399 & $21.11\pm0.06$ & 3 & $20.04\pm0.20$ & 1 & $21.18\pm0.06$ & $0.26$ & $0.15$ \\
HD 116781 & $21.18\pm0.10$ & 1 & $20.08\pm0.05$ & 2 & $21.24\pm0.09$ & \nodata & $0.14$ \\
HD 122879 & $21.26\pm0.12$ & 4 & $20.24\pm0.09$ & 4 & $21.34\pm0.10$ & $0.15$ & $0.16$ \\
HD 124314 & $21.34\pm0.10$ & 3 & $20.47\pm0.03$ & 2 & $21.44\pm0.08$ & $0.82$ & $0.21$ \\
HD 147888 & $21.71\pm0.09$ & 4 & $20.47\pm0.05$ & 6 & $21.76\pm0.08$ & $13.63$ & $0.10$ \\
HD 152590 & $21.37\pm0.06$ & 4 & $20.47\pm0.07$ & 4 & $21.47\pm0.05$ & $0.53$ & $0.20$ \\
HD 168941 & $21.11\pm0.09$ & 3 & $20.10\pm0.05$ & 2 & $21.19^{+0.08}_{-0.07}$ & $0.10$ & $0.16$ \\
HD 177989 & $20.95\pm0.09$ & 3 & $20.12\pm0.05$ & 2 & $21.06\pm0.07$ & $0.07$ & $0.23$ \\
HD 185418 & $21.15\pm0.09$ & 8 & $20.76\pm0.05$ & 5 & $21.41^{+0.06}_{-0.05}$ & $0.87$ & $0.45$ \\
HD 192639 & $21.29\pm0.08$ & 8 & $20.69\pm0.05$ & 5 & $21.47^{+0.06}_{-0.05}$ & $0.86$ & $0.33$ \\
HD 195965 & $20.95\pm0.03$ & 10 & $20.37\pm0.03$ & 2 & $21.13\pm0.02$ & $0.34$ & $0.34$ \\
HD 202347 & $20.99\pm0.10$ & 1 & $20.00\pm0.06$ & 9 & $21.07^{+0.09}_{-0.08}$ & $0.29$ & $0.17$ \\
HD 203374 & $21.11\pm0.09$ & 11 & $20.68\pm0.05$ & 2 & $21.35^{+0.06}_{-0.05}$ & $0.89$ & $0.43$ \\
HD 206267 & $21.30\pm0.15$ & 5 & $20.86\pm0.04$ & 5 & $21.54^{+0.09}_{-0.08}$ & $1.12$ & $0.42$ \\
HD 207198 & $21.34\pm0.17$ & 3 & $20.83\pm0.04$ & 5 & $21.55^{+0.11}_{-0.10}$ & $1.15$ & $0.38$ \\
HD 207308 & $21.20\pm0.10$ & 1 & $20.76\pm0.05$ & 2 & $21.44\pm0.06$ & \nodata & $0.42$ \\
HD 207538 & $21.34\pm0.12$ & 3 & $20.91\pm0.06$ & 5 & $21.58^{+0.08}_{-0.07}$ & $1.40$ & $0.43$ \\
HD 209339 & $21.16\pm0.10$ & 1 & $20.19\pm0.03$ & 2 & $21.24\pm0.08$ & $0.52$ & $0.18$ \\
HD 210839 & $21.19\pm0.06$ & 8 & $20.84\pm0.04$ & 5 & $21.47\pm0.04$ & $1.88$ & $0.47$ \\
HD 224151 & $21.32\pm0.08$ & 8 & $20.57\pm0.05$ & 2 & $21.45\pm0.06$ & $0.68$ & $0.26$ \\
HD 303308 & $21.45\pm0.08$ & 3 & $20.23\pm0.05$ & 2 & $21.50\pm0.07$ & $0.39$ & $0.11$ \\
\enddata
\tablerefs{(1) \citet{JensenFeII}.  (2) J.~M.~Shull et al., in preparation.  (3) \citet{DS1994}.  (4) \citet{Cartledge2004}.  (5) \citet{Rachford2002}.  (6) B.~L.~Rachford et al., in preparation.  (7) \citet{JensenNI}.  (8) \citet{Andre2003}.  (9) \citet{SDJ2007}.  (10) \citet{Hoopes2003}.  (11) \citet{DS1994}, Table 2 (lines of sight with uncertain stellar parameters).}
\end{deluxetable}

\clearpage \clearpage

\begin{deluxetable}{ccccccc}
\tablecolumns{7}
\tablewidth{0pc}
\tabletypesize{\tiny}
\tablecaption{Lines of Sight: Reddening Data\label{MgII_reddeningtable}}
\tablehead{\colhead{Line of Sight} & \colhead{$\ebv$} & \colhead{Ref.} & \colhead{$\av$} & \colhead{Ref.} & \colhead{$\rv$} & \colhead{Ref.} \\ \colhead{} & \colhead{(mag)} & \colhead{} & \colhead{(mag)} & \colhead{} & \colhead{} & \colhead{}}
\startdata
BD +35$^{\circ}$4258 & 0.29 & 1 & 0.93 & 2 & 3.21 & 3 \\
CPD -59$^{\circ}$2603 & 0.46 & 4 & 1.45 & 3 & 3.15 & 5 \\
HD 12323 & 0.24 & 1 & 0.74 & 2 & 3.08 & 3 \\
HD 13745 & 0.46 & 4 & 1.42 & 2 & 3.09 & 3 \\
HD 15137 & 0.31 & 4 & 1.10 & 2 & 3.55 & 3 \\
HD 27778 & 0.37 & 6 & 1.01 & 3 & 2.73 & 6 \\
HD 37021 & 0.54 & 6 & 2.99 & 3 & 5.54 & 6 \\
HD 37061 & 0.52 & 6 & 2.20 & 3 & 4.23 & 6 \\
HD 37903 & 0.35 & 4 & 1.28 & 3 & 3.66 & 6 \\
HD 40893 & 0.46 & 6 & 1.13 & 3 & 2.46 & 6 \\
HD 66788 & 0.20 & 5 & 0.69 & 2 & 3.45 & 3 \\
HD 69106 & 0.18 & 6 & 0.58 & 2 & 3.22 & 3 \\
HD 91597 & 0.27 & 6 & 1.33 & 2 & 4.93 & 3 \\
HD 91651 & 0.30 & 4 & 0.95 & 2 & 3.17 & 3 \\
HD 92554 & 0.39 & 4 & 1.15 & 2 & 2.95 & 3 \\
HD 93205 & 0.37 & 4 & 1.21 & 2 & 3.27 & 3 \\
HD 93222 & 0.37 & 7 & 1.18 & 2 & 3.19 & 3 \\
HD 93843 & 0.34 & 4 & 1.03 & 2 & 3.03 & 3 \\
HD 94493 & 0.20 & 4 & 0.71 & 2 & 3.55 & 3 \\
HD 99857 & 0.33 & 4 & 1.10 & 2 & 3.33 & 3 \\
HD 99890 & 0.24 & 4 & 0.72 & 2 & 3.00 & 3 \\
HD 103779 & 0.21 & 4 & 0.68 & 2 & 3.24 & 3 \\
HD 104705 & 0.26 & 4 & 0.80 & 2 & 3.08 & 3 \\
HD 109399 & 0.26 & 4 & 0.81 & 2 & 3.12 & 3 \\
HD 116781 & 0.34 & 5 & 1.40 & 2 & 4.12 & 3 \\
HD 122879 & 0.36 & 7 & 1.12 & 2 & 3.11 & 3 \\
HD 124314 & 0.53 & 4 & 1.64 & 2 & 3.09 & 3 \\
HD 147888 & 0.47 & 6 & 1.91 & 3 & 4.06 & 6 \\
HD 152590 & 0.46 & 6 & 1.51 & 3 & 3.28 & 6 \\
HD 168941 & 0.37 & 4 & 1.03 & 2 & 2.78 & 3 \\
HD 177989 & 0.25 & 4 & 0.25 & 2 & 1.00 & 3 \\
HD 185418 & 0.50 & 6 & 1.16 & 3 & 2.32 & 6 \\
HD 192639 & 0.66 & 4 & 1.87 & 3 & 2.83 & 6 \\
HD 195965 & 0.25 & 4 & 0.77 & 2 & 3.08 & 3 \\
HD 202347 & 0.19 & 8 & 0.53 & 2 & 2.79 & 3 \\
HD 203374 & 0.60 & 9 & 1.88 & 2 & 3.13 & 3 \\
HD 206267 & 0.53 & 6 & 1.41 & 3 & 2.66 & 6 \\
HD 207198 & 0.62 & 6 & 1.50 & 3 & 2.42 & 6 \\
HD 207308 & 0.52 & 8 & 1.61 & 2 & 3.10 & 3 \\
HD 207538 & 0.64 & 6 & 1.44 & 3 & 2.25 & 6 \\
HD 209339 & 0.38 & 8 & 1.09 & 2 & 2.87 & 3 \\
HD 210839 & 0.57 & 4 & 1.58 & 3 & 2.77 & 6 \\
HD 224151 & 0.44 & 4 & 1.52 & 2 & 3.45 & 3 \\
HD 303308 & 0.45 & 7 & 1.09 & 2 & 2.42 & 3 \\
\enddata
\tablerefs{(1) \citet{Savage1985}.  (2) \citet{Neckel1980}.  (3) Derived from the other two quantities via the relationship $\rv \equiv \av / \ebv$.  (4) \citet{DS1994}.  (5) \citet{JensenFeII}.  (6) DIB Database, http://dib.uiuc.edu; $\ebv$ compiled by L.~M.~Hobbs, $\rv$ derived by B.~L.~Rachford through polarization or infrared photometry \citet[for method and examples, see][]{Rachford2002}.  (7) \citet{Aiello1988}.  (8) \citet{Garmany1992}.  (9) \citet{DS1994}, Table 2 (lines of sight with uncertain stellar parameters).}
\end{deluxetable}

\clearpage \clearpage

\begin{deluxetable}{cccccc}
\tablecolumns{6}
\tablewidth{0pc}
\tabletypesize{\tiny}
\tablecaption{Mg II Column Density Results\label{MgII_coldensities}}
\tablehead{\colhead{Line of Sight} & \colhead{$\logMgII$} & \colhead{$\logMgIIH$} & \colhead{Mg/H (ppm)} & \colhead{$W_{1239.9}$} & \colhead{$W_{1240.4}$}}
\startdata
BD +35$^{\circ}$4258 & $16.15^{+0.06}_{-0.08}$ & $-5.15^{+0.10}_{-0.15}$ & $7.1^{+1.8}_{-2.1}$ & $81.4\pm10.4$ & $54.8\pm10.4$ \\
CPD -59$^{\circ}$2603 & $16.35\pm0.03$ & $-5.15^{+0.06}_{-0.08}$ & $7.1^{+1.1}_{-1.3}$ & $112.4\pm6.6$ & $78.8\pm7.4$ \\
HD 12323 & $16.06^{+0.04}_{-0.05}$ & $-5.23^{+0.07}_{-0.10}$ & $5.9^{+1.1}_{-1.3}$ & $64.3\pm5.1$ & $43.9\pm6.0$ \\
HD 13745 & $16.18^{+0.06}_{-0.07}$ & $-5.20^{+0.08}_{-0.12}$ & $6.3^{+1.3}_{-1.5}$ & $93.3\pm10.1$ & $61.3\pm10.9$ \\
HD 15137 & $16.15\pm0.02$ & $-5.07^{+0.09}_{-0.19}$ & $8.5^{+2.1}_{-3.0}$ & $72.1\pm2.4$ & $50.9\pm2.9$ \\
HD 27778 & $15.42^{+0.05}_{-0.06}$ & $-5.98^{+0.07}_{-0.11}$ & $1.1\pm0.2$ & $16.0\pm1.7$ & $10.6\pm1.5$ \\
HD 37021 & $15.93^{+0.03}_{-0.04}$ & $-5.75^{+0.10}_{-0.17}$ & $1.8^{+0.5}_{-0.6}$ & $33.7\pm2.2$ & $23.7\pm2.2$ \\
HD 37061 & $15.78\pm0.03$ & $-5.95^{+0.08}_{-0.12}$ & $1.1^{+0.2}_{-0.3}$ & $27.0\pm1.5$ & $19.8\pm1.4$ \\
HD 37903 & $15.55^{+0.07}_{-0.09}$ & $-5.94^{+0.09}_{-0.12}$ & $1.1\pm0.3$ & $21.1\pm3.1$ & $14.2\pm3.2$ \\
HD 40893 & $16.33\pm0.03$ & $-5.26^{+0.07}_{-0.11}$ & $5.5^{+1.0}_{-1.2}$ & $95.5\pm5.0$ & $69.8\pm5.3$ \\
HD 66788 & $16.11^{+0.08}_{-0.10}$ & $-5.15^{+0.11}_{-0.17}$ & $7.1^{+2.0}_{-2.3}$ & $81.2\pm10.6$ & $39.5\pm11.1$ \\
HD 69106 & $15.81^{+0.02}_{-0.03}$ & $-5.30^{+0.05}_{-0.07}$ & $5.0\pm0.7$ & $37.3\pm1.7$ & $25.3\pm1.8$ \\
HD 91597 & $16.25\pm0.05$ & $-5.16^{+0.07}_{-0.09}$ & $6.8^{+1.2}_{-1.3}$ & $94.4\pm8.7$ & $66.1\pm9.6$ \\
HD 91651 & $16.26^{+0.03}_{-0.04}$ & $-4.90^{+0.06}_{-0.08}$ & $12.7^{+1.9}_{-2.1}$ & $108.0\pm6.6$ & $72.3\pm6.9$ \\
HD 92554 & $16.37^{+0.04}_{-0.05}$ & $-4.91^{+0.09}_{-0.14}$ & $12.3^{+2.8}_{-3.4}$ & $109.4\pm9.1$ & $80.0\pm9.3$ \\
HD 93205 & $16.32^{+0.02}_{-0.03}$ & $-5.08^{+0.05}_{-0.06}$ & $8.3^{+1.0}_{-1.1}$ & $113.7\pm4.9$ & $77.9\pm5.9$ \\
HD 93222 & $16.41^{+0.01}_{-0.02}$ & $-5.01^{+0.06}_{-0.08}$ & $9.8^{+1.4}_{-1.7}$ & $114.4\pm3.0$ & $84.3\pm3.6$ \\
HD 93843 & $16.25\pm0.02$ & $-5.10^{+0.07}_{-0.10}$ & $7.9^{+1.3}_{-1.6}$ & $104.0\pm3.4$ & $69.8\pm4.1$ \\
HD 94493 & $16.16\pm0.02$ & $-5.00^{+0.04}_{-0.05}$ & $9.9\pm1.1$ & $77.2\pm3.2$ & $53.5\pm3.6$ \\
HD 99857 & $16.21\pm0.03$ & $-5.11^{+0.06}_{-0.09}$ & $7.7^{+1.2}_{-1.4}$ & $80.6\pm4.1$ & $55.8\pm4.4$ \\
HD 99890 & $16.18\pm0.03$ & $-4.78^{+0.10}_{-0.18}$ & $16.6^{+4.2}_{-5.6}$ & $83.4\pm4.5$ & $57.3\pm5.3$ \\
HD 103779 & $16.17\pm0.02$ & $-5.02^{+0.08}_{-0.12}$ & $9.5^{+1.9}_{-2.3}$ & $81.6\pm3.4$ & $56.0\pm3.9$ \\
HD 104705 & $16.19\pm0.02$ & $-4.98^{+0.06}_{-0.08}$ & $10.4^{+1.4}_{-1.7}$ & $83.8\pm3.2$ & $57.2\pm3.4$ \\
HD 109399 & $15.95\pm0.05$ & $-5.23^{+0.07}_{-0.09}$ & $5.9^{+1.0}_{-1.1}$ & $56.8\pm5.0$ & $36.9\pm5.4$ \\
HD 116781 & $16.14\pm0.03$ & $-5.11^{+0.08}_{-0.12}$ & $7.8^{+1.5}_{-1.8}$ & $75.5\pm4.4$ & $51.8\pm4.8$ \\
HD 122879 & $16.22\pm0.02$ & $-5.11^{+0.08}_{-0.14}$ & $7.7^{+1.6}_{-2.1}$ & $86.5\pm2.8$ & $60.4\pm3.2$ \\
HD 124314 & $16.32\pm0.02$ & $-5.12^{+0.07}_{-0.10}$ & $7.6^{+1.3}_{-1.6}$ & $97.6\pm3.0$ & $70.9\pm3.7$ \\
HD 147888 & $15.83\pm0.03$ & $-5.93^{+0.07}_{-0.11}$ & $1.2^{+0.2}_{-0.3}$ & $24.1\pm1.5$ & $18.9\pm1.5$ \\
HD 152590 & $16.20\pm0.03$ & $-5.26^{+0.05}_{-0.07}$ & $5.5^{+0.7}_{-0.8}$ & $71.3\pm4.3$ & $52.1\pm4.2$ \\
HD 168941 & $15.87^{+0.08}_{-0.10}$ & $-5.32^{+0.10}_{-0.14}$ & $4.8^{+1.2}_{-1.3}$ & $42.8\pm7.0$ & $28.9\pm7.0$ \\
HD 177989 & $15.83\pm0.03$ & $-5.23^{+0.06}_{-0.09}$ & $5.9^{+0.9}_{-1.1}$ & $40.7\pm2.0$ & $27.1\pm2.2$ \\
HD 185418 & $15.94\pm0.03$ & $-5.47^{+0.05}_{-0.07}$ & $3.4^{+0.4}_{-0.5}$ & $41.7\pm2.1$ & $30.0\pm2.1$ \\
HD 192639 & $16.20\pm0.03$ & $-5.26^{+0.05}_{-0.07}$ & $5.4^{+0.7}_{-0.8}$ & $60.3\pm3.2$ & $46.9\pm3.7$ \\
HD 195965 & $15.89^{+0.04}_{-0.05}$ & $-5.24\pm0.05$ & $5.7\pm0.7$ & $45.0\pm3.4$ & $30.3\pm4.1$ \\
HD 202347 & $15.62^{+0.04}_{-0.05}$ & $-5.45^{+0.08}_{-0.12}$ & $3.5^{+0.7}_{-0.8}$ & $27.5\pm2.2$ & $17.6\pm2.2$ \\
HD 203374 & $16.07\pm0.02$ & $-5.28^{+0.05}_{-0.07}$ & $5.3^{+0.7}_{-0.8}$ & $63.3\pm2.7$ & $44.3\pm2.8$ \\
HD 206267 & $16.05\pm0.04$ & $-5.48^{+0.08}_{-0.13}$ & $3.3^{+0.6}_{-0.9}$ & $58.1\pm4.2$ & $40.7\pm4.7$ \\
HD 207198 & $16.08\pm0.02$ & $-5.47^{+0.08}_{-0.16}$ & $3.4^{+0.7}_{-1.0}$ & $56.6\pm2.2$ & $40.8\pm2.7$ \\
HD 207308 & $15.93\pm0.05$ & $-5.51^{+0.07}_{-0.10}$ & $3.1^{+0.5}_{-0.6}$ & $45.9\pm4.1$ & $31.7\pm4.6$ \\
HD 207538 & $16.07^{+0.03}_{-0.04}$ & $-5.51^{+0.07}_{-0.10}$ & $3.1^{+0.5}_{-0.6}$ & $58.2\pm3.5$ & $41.4\pm4.1$ \\
HD 209339 & $16.04\pm0.02$ & $-5.21^{+0.07}_{-0.11}$ & $6.2^{+1.1}_{-1.3}$ & $53.9\pm1.8$ & $38.4\pm2.0$ \\
HD 210839 & $16.05\pm0.03$ & $-5.42^{+0.04}_{-0.05}$ & $3.8\pm0.4$ & $63.1\pm3.3$ & $42.9\pm3.7$ \\
HD 224151 & $16.30\pm0.03$ & $-5.15^{+0.06}_{-0.08}$ & $7.1^{+1.0}_{-1.2}$ & $104.6\pm5.8$ & $72.4\pm6.5$ \\
HD 303308 & $16.34^{+0.04}_{-0.05}$ & $-5.15^{+0.07}_{-0.10}$ & $7.0^{+1.3}_{-1.4}$ & $108.1\pm8.6$ & $75.8\pm9.0$ \\
\enddata
\end{deluxetable}

\clearpage \clearpage

\begin{deluxetable}{ccccc}
\tablecolumns{5}
\tablewidth{0pc}
\tabletypesize{\scriptsize}
\tablecaption{Comparison with CLMS\label{MgII_comparison}}
\tablehead{\colhead{Line of Sight} & \multicolumn{4}{c}{$\logMgII$} \\ \colhead{} & \colhead{This paper} & \colhead{CLMS AOD\tablenotemark{a}} & \colhead{CLMS PF\tablenotemark{b}} & \colhead{CLMS AE\tablenotemark{c}}}
\startdata
HD 12323 & $16.06^{+0.04}_{-0.05}$ & $16.03\pm0.02$ & $16.04\pm0.05$ & $0.05$ \\
HD 27778 & $15.42^{+0.05}_{-0.06}$ & $15.49\pm0.02$ & $15.48\pm0.01$ & $0.02$ \\
HD 37021 & $15.93^{+0.03}_{-0.04}$ & $15.85\pm0.02$ & $15.90\pm0.02$ & $0.03$ \\
HD 37061 & $15.78\pm0.03$ & $15.78\pm0.01$ & $15.80\pm0.05$ & $0.05$ \\
HD 37903 & $15.55^{+0.07}_{-0.09}$ & $15.65\pm0.02$ & $15.62\pm0.06$ & $0.06$ \\
HD 122879 & $16.22\pm0.02$ & $16.22\pm0.02$ & $16.23\pm0.01$ & $0.02$ \\
HD 147888 & $15.83\pm0.03$ & $15.84\pm0.03$ & $16.03\pm0.03$ & $0.04$ \\
HD 152590 & $16.20\pm0.03$ & $16.19\pm0.02$ & $16.20\pm0.09$ & $0.09$ \\
HD 185418 & $15.94\pm0.03$ & $15.97\pm0.02$ & $15.96\pm0.01$ & $0.02$ \\
HD 192639 & $16.20\pm0.03$ & $16.19\pm0.02$ & $16.21\pm0.02$ & $0.03$ \\
HD 207198 & $16.08\pm0.02$ & $16.08\pm0.01$ & $16.08\pm0.02$ & $0.02$ \\
\enddata
\tablenotetext{a}{CLMS results, apparent optical depth method (uncorrected).}
\tablenotetext{b}{CLMS results, profile fitting method.}
\tablenotetext{c}{Adopted error of CLMS.  Note that CLMS adopt profile fitting results for the best column density, but adopt errors that are the errors of the apparent optical depth and profile fitting methods added in quadrature.}
\end{deluxetable}

\clearpage \clearpage

\clearpage \clearpage

\clearpage \clearpage

\clearpage \clearpage

\clearpage \clearpage

\clearpage \clearpage

\end{document}